\documentclass[superscriptaddress,twocolumn,showpacs,floatfix,prc,nofootinbib]{revtex4}
\usepackage{dcolumn}
\usepackage{bm}
\usepackage{color}

\usepackage{graphicx}
\usepackage{amsmath}
\usepackage{amssymb}
\usepackage{slashed}
\usepackage{tabularx}
\usepackage{wrapfig}
\def\beq{\begin{equation}}
\def\eeq{\end{equation}}
\def\be{\begin{eqnarray}}
\def\ee{\end{eqnarray}}

\newcolumntype{L}[1]{>{\hsize=#1\hsize\raggedright\arraybackslash}X}%
\newcolumntype{R}[1]{>{\hsize=#1\hsize\raggedleft\arraybackslash}X}%
\newcolumntype{C}[1]{>{\hsize=#1\hsize\centering\arraybackslash}X}%

\begin{document}
\title{Weak production of  strange and charm ground-state baryons in nuclei}
\author{J. E. Sobczyk}
\affiliation{Instituto de F\'\i sica Corpuscular (IFIC), Centro Mixto
CSIC-Universidad de Valencia, Institutos de Investigaci\'on de
Paterna, Apartado 22085, E-46071 Valencia, Spain}
\author{N. Rocco}
\affiliation{Physics Division, Argonne National Laboratory, Argonne, Illinois 60439, USA}
\affiliation{Theoretical Physics Department, Fermi National Accelerator Laboratory, P.O. Box 500, Batavia, IL 60510, USA}
\author{A. Lovato}
\affiliation{Physics Division, Argonne National Laboratory, Argonne, Illinois 60439, USA}
\affiliation{INFN-TIFPA Trento Institute of Fundamental Physics and Applications, 38123 Trento, Italy}
\author{J. Nieves}
\affiliation{Instituto de F\'\i sica Corpuscular (IFIC), Centro Mixto
CSIC-Universidad de Valencia, Institutos de Investigaci\'on de
Paterna, Apartado 22085, E-46071 Valencia, Spain}

\date{\today}
\begin{abstract}

We present results for the quasi-elastic weak production of $\Lambda$ and $\Sigma$ hyperons induced by $\bar{\nu}$
scattering off nuclei, in the kinematical region of interest for accelerator neutrino experiments. 
We employ realistic hole spectral functions and we describe the propagation of the hyperons in the nuclear medium 
by means of a Monte Carlo cascade. The latter strongly modifies the kinematics and the relative production rates 
of the hyperons, leading to a non-vanishing $\Sigma^+$ cross section, to a sizable enhancement of the 
$\Lambda$ production and to a drastic reduction of the $\Sigma^0$ and $\Sigma^-$ distributions.  
We also compute the quasi-elastic weak $\Lambda_c$ production cross section, paying special attention
to estimate the uncertainties induced by the model dependence of the vacuum $n\to \Lambda_c$ weak matrix element.
In this regard, the recent BESIII measurements of the branching ratios of $\Lambda_c\rightarrow \Lambda l^+\nu_l$
($l=e,\mu$) are used to benchmark the available theoretical predictions. 

\end{abstract}
\maketitle

\section{Introduction}

In this work we analyze the quasi-elastic (QE) weak production of $\Lambda$ and $\Sigma$ hyperons induced by $\bar{\nu}$ scattering off nuclei. Electroweak charged-current can induce the production of strange particles through both the $\Delta S=0$ and $\Delta S=1$ channels, where $S$ denotes the strangeness quantum number. We focus our analysis on $\Delta S=1$ processes, for initial $\bar{\nu}$ energies in the range $E_\nu=1-3$ GeV, of interest for accelerator neutrino experiments. Although $\Delta S=1$ transitions are generally suppressed with respect to $\Delta S=0$ reactions by a factor $\tan\theta_C$ -- $\theta_C$ being the Cabibbo angle -- in this energy region the latter is strongly reduced by the available phase space and the rates of two processes become comparable. To describe the propagation of hyperons in the nuclear medium we have devised a Monte Carlo cascade (MCC) algorithm. We treat the rescattering processes that the hyperon undergoes before exiting the nucleus in a classical way, using the experimental data for the hyperon-nucleon cross sections as an input. 
 
Our analysis improves upon the pioneering work of Ref.~\cite{Singh:2006xp}, where nuclear effects were for the first time included in the description of $\Delta S=1$ processes, although a simplified description of the initial nuclear target -- the Local Fermi gas (LFG) model --  was employed. In this work, we provide a realistic description of the nuclear structure, employing two distinct hole spectral functions (SFs). We analyze the role of nuclear correlations in the nuclear ground-state and final state interactions (FSI) between the produced hyperon and the spectator nucleons, in double and single-differential, and total inclusive $\bar{\nu}_l+^{16}$O cross sections in which $\Lambda$ and $\Sigma$ hyperons are produced. The dependence of these effects on the initial nuclear species is also discussed by comparing the results obtained for total cross sections in $^{12}$C,$^{16}$O, and $^{40}$Ca. All these findings might have important implications in the analysis of SciBooNE~\cite{sciboone_web}, MicroBooNE~\cite{microboone_web}, MINERvA~\cite{minerva_web} and ArgoNeuT~\cite{argoneut_web} experiments once the data collected using $\bar{\nu}$ beams become available.

The BESIII Collaboration has recently reported on the absolute measurement of the branching ratios of $\Lambda_c\rightarrow \Lambda e^+\nu_e$~\cite{Ablikim:2015prg} and $\Lambda_c\rightarrow \Lambda \mu^+\nu_\mu$~\cite{Ablikim:2016vqd}, which can serve as an important benchmark to compare various theoretical predictions. These decays have been studied within several relativistic and nonrelativistic constituent quark models (CQMs), which provide predictions for the transition form factors~\cite{Hussain:2017lir, Gutsche:2015rrt, PerezMarcial:1989yh,Faustov:2016yza}. Some of the groups have also presented results for the $\Lambda_c\rightarrow N$ form factors~\cite{Gutsche:2014zna, PerezMarcial:1989yh}.  The first lattice QCD (LQCD) calculation of the $\Lambda_c \to \Lambda$ and $\Lambda_c \to N$  form factors reported in Refs.~\cite{Meinel:2016dqj} and  \cite{Meinel:2017ggx}, respectively, deserves special mention. 
 
Capitalizing on these recent developments, in this work we also present realistic results for the weak $\Lambda_c$ production cross section in nuclei, providing estimates of the theoretical uncertainty of our predictions. We limit our analysis to relatively-low neutrino energies, close to threshold, and we assume that the QE mechanism, $W^+ n_b\to \Lambda_c$ where $n_b$ is a neutron bound in the nucleus, is the dominant one. To estimate the theoretical uncertainties related to the elementary amplitudes, we consider two different scenarios. In the first one, we relate the $\Lambda_c\rightarrow \Lambda$ and $\Lambda_c\rightarrow N$ form factors by a SU(3) rotation, neglecting the effects of SU(3) symmetry breaking. In the second one, we directly use $\Lambda_c\rightarrow N$ form factors from theoretical calculations, whenever available. 

The kinematical region of interest for neutrino experiments corresponds to $q^2={\vec{q}}\,^2-(q^0)\,^2\in(-5,0)$ GeV$^2$, where $\vec{q}$ and $q^0$ are the energy and momentum transfers of the scattering process. Since the semileptonic $\Lambda_c$ decay  occurs for $q^2>0$, the analysis of the neutrino cross section requires an extrapolation of the form factors to negative $q^2$, which is not expected to be completely free of uncertainties.  Therefore, in our analysis, we select models that are capable to reproduce the measured ${\cal B}(\Lambda_c\rightarrow \Lambda e^+\nu_e)$ branching ratio, and whose form factor parameterizations do not lead to pathological behaviors in the $q^2<0$ region. We use four different models for the $\Lambda_c\rightarrow N$ form factors to compute the $\nu_\mu+\, ^{16}{\rm O} \to \mu^- \Lambda_c X$ total cross section. Particular emphasis shall be devoted to the results obtained employing the analytical continuation of the LQCD calculations of Ref.~\cite{Meinel:2017ggx}. The latter are supplemented by estimates of the theoretical uncertainties that we propagate throughout our calculations to the neutrino cross sections. A more conservative theoretical uncertainty is obtained from the spread of the predictions found when the different sets of CQM form factors are also considered.

Finally, we would like to point out that since the calculation of the total nuclear cross section and the semileptonic $\Lambda_c\to \Lambda$ decay width are sensitive to different kinematical regions, and therefore presumably to different form factors, the combined study of the two processes allows to better elucidate the differences between models.

This work is organized as follows. In Sec.~\ref{sec:form}, we provide the general expressions for hyperon production cross section in the vacuum and the definition of the form factors used to parameterize the weak matrix elements. In Subsec.~\ref{sec:lambdaC}, we outline the main features of the different schemes examined in the case of the $N\to\Lambda_c$  transition. Our approach to account for SF and FSI nuclear effects is described in Sec.~\ref{sec:sfs}, while the main results of this study are presented in Sec.~\ref{sec:results}. Finally, we collect the main conclusions of this work in Sec.~\ref{sec:concl}, while we provide further details on the $\Lambda_c\to \Lambda, N$ form factors in the Appendix. 

\section{Formalism}
\label{sec:form}
\subsection{Cross section}
\label{nu:nucl:scatt}

The unpolarized differential cross section for the $\bar{\nu}_l(k) + N(p) \rightarrow l^+(k')+Y(p')$ reaction, in which an antineutrino $\bar{\nu}_l$ scatters off a nucleon $N$, and in the final state the charged lepton $l^+$ and the hyperon $Y$ are produced, is given in the laboratory frame by\footnote{In the case of $\Lambda_c$ production, the reaction is obviously induced by neutrinos instead of antineutrinos. We will come back to this point below.} 
\begin{equation}
\frac{d\sigma}{dE_{k'}d\Omega(\hat{k}')}=\frac{G_F^2 \sin^2\theta_C}{4\pi^2}\frac{|\vec{k}'|}{|\vec{k}\,|} L_{\mu\sigma}^{(\bar\nu)}(k,k')W^{\mu\sigma}_N(p,q)\, .
\label{eq:sigma2D}
\end{equation}
In the above equation $G_F=1.1664\times 10^{-5}$ GeV$^{-2}$ is the Fermi constant, $\theta_C$ the Cabibbo angle ($\sin\theta_C=0.225$), $p^\mu = (M,\vec{0}\,)$, with $M$ the nucleon mass,  $q=k-k'$. The lepton and nucleon tensors read
\begin{align}
L_{\mu\sigma}^{(\bar\nu)}(k,k')  & = k_\mu k'_\sigma+k'_\mu k_\sigma-g_{\mu\sigma}k\cdot k'-i\epsilon_{\mu\sigma\alpha\beta}k^{\prime\alpha} k^\beta \nonumber\\
W^{\mu\sigma}_N(p,q)&=2M_Y\int\frac{d^3p'}{2E_{p'}^Y}\delta^4(q+p-p')  A^{\mu\sigma}(p,q,p') \nonumber\\
&= \frac{M_Y}{ E_{p'}^Y}\delta(q^0+M-E_{p'}^Y)A^{\mu\sigma}(p,q,p+q) 
\label{eq:phasespce}
\end{align}
with $\epsilon_{0123}=+1$, the metric $g^{\mu\nu}=(+,-,-,-)$, and 
\begin{equation}
A^{\mu\sigma}(p,q)= \frac12 {\rm Tr}\left[ \frac{(/\hspace{-.2cm} p+/\hspace{-.2cm} q+ M_Y)}{2M_Y} \Gamma^\mu_Y \frac{(/\hspace{-.2cm} p+M)}{2M}\gamma^0\Gamma^{\sigma\dagger}_Y\gamma^0\right] \label{eq:amunu}
\end{equation}
where $M_Y$ is the hyperon mass. The Dirac operator $\Gamma^\mu_Y$ is determined from the hadronic matrix element of the charged-current operator (we use spinors normalized as $\bar u u = 1$), 
\begin{equation}
J^\mu = \langle Y(p') | V^\mu-A^\mu| N(p)\rangle = \bar{u}_Y(p') \Gamma^\mu_Y u(p)
\label{jmu:1n} 
\end{equation}
where
\begin{align}
\Gamma^\mu_Y &= \big[ \gamma^\mu f_1(q^2)+ i\sigma^{\mu\nu} \frac{q_\nu}{M_Y} f_2(q^2)+ \frac{q^\mu}{M_Y} f_3(q^2) \big] \nonumber \\
& -\big[ \gamma^\mu g_1(q^2)+ i\sigma^{\mu\nu} \frac{q_\nu}{M_Y} g_2(q^2)+ \frac{q^\mu}{M_Y} g_3(q^2) \big] \gamma_5
\label{eq:defffs}
\end{align}
and dimensionless axial and vector transition form factors, $f_i(q^2)$ and $g_i(q^2)$, describe the $N\rightarrow Y$ weak transition.

The integration over $E_{k'}$ needed to obtain the angular cross section can be carried out exploiting the energy-conserving delta function of Eq.~(\ref{eq:phasespce})
\begin{align}
I(|\vec{k}\,|,\cos\theta')&= \int dE_{k'} |\vec{k}'| \frac{\delta\left(q^0+M-\sqrt{M_Y^2+\vec{q}^{\, 2}}\right)}{\sqrt{M_Y^2+\vec{q}^{\, 2}}}\nonumber\\
&= \frac{|\vec{k}'|}{M +|\vec{k}\,|-\frac{E_{k'}}{|\vec{k}'|}|\vec{k}\,|\cos\theta'}
\end{align}
where $|\vec{k}'|$ depends on the lepton scattering angle $\theta'$ through the energy conservation equation $E_{k'}=|\vec{k}\,|+M-\sqrt{M_Y^2+\vec{q}^{\, 2}}$, with $\vec{q}^{\, 2} = \vec{k}^2+\vec{k}^{'2}- 2|\vec{k}\,||\vec{k}'|\cos\theta'$, $E_{k'}= \sqrt{\vec{k}^{'2}+m^2_l}$ and $m_l$ is the mass of the outgoing lepton. Neglecting $m_l$, the expressions simplify to
\begin{align}
 \frac{d\sigma}{d\Omega(\hat{k}')}&=\frac{G_F^2 \sin^2\theta_C}{4\pi^2}\,\frac{M_Y}{M}\,\frac{|\vec{k'}|^2}{|\vec{k}\,|^2}\,
 \frac{L_{\mu\sigma}^{(\bar\nu)}(k,k')A^{\mu\sigma}(p,q)}{1+\frac{M^2-M^2_Y}{2|\vec{k}\,|M}},
 \end{align}
 with
\begin{equation}
E_{k'} = |\vec{k}'|= \frac{|\vec{k}\,|M+(M^2-M^2_Y)/2}{M+|\vec{k}\,|(1-\cos\theta')}
\end{equation}

The differential cross section for the neutrino $\nu_l(k) + N(p) \rightarrow l^-(k')+Y_C(p')$ charm production reaction can be obtained from the above expressions replacing $L_{\mu\sigma}^{(\bar\nu)}$ by $L_{\mu\sigma}^{\nu} = L_{\sigma\mu}^{\bar\nu}$ and using the appropriate masses and form factors.


\begin{table*}[t]
\begin{center}
\begin{tabularx}{0.8\textwidth}{ C{0.4} | C{1.2} C{1.4} C{1} }
\hline
\hline 
\\
  & $f_1(q^2)$ & $f_2(q^2)$ &$g_1(q^2)$ \\ \\
  \hline
  \\
  $p\rightarrow \Lambda$ & $-\sqrt{\frac{3}{2}}f_1^p(q^2)$ & $-\sqrt{\frac{3}{2}}\frac{M_Y}{M+M_Y}f_2^p(q^2)$ & $-\sqrt{\frac{3}{2}}\frac{1+2x}{3}g_A(q^2)$ \\ \\
    $n\rightarrow \Sigma^-$  & $-\left(f_1^p(q^2)+2f_1^n(q^2)\right)$ & $-\frac{M_Y}{M+M_Y}(f_2^p(q^2)+2f_2^n(q^2))$ & $(1-2x) g_A(q^2)$ \\ \\
    $p\rightarrow \Sigma^0$  & $-\frac{1}{\sqrt{2}}\left(f_1^p(q^2)+2f_1^n(q^2)\right)$ & $-\frac{M_Y}{M+M_Y}\frac{1}{\sqrt{2}}\left(f_2^p(q^2)+2f_2^n(q^2)\right)$ & $\frac{1-2x}{\sqrt{2}} g_A(q^2)$ \\ \\

  \hline
  \hline  
\end{tabularx}
\end{center}
\caption{Form factors for $\Delta S=1$ transitions, with $x\approx 0.73$.}
\label{table:ff}
\end{table*}

\subsection{Form factors}
\subsubsection{$N \rightarrow \Lambda,\Sigma^0,\Sigma^-$ transitions}

The parametrization adopted for the form factors of the $N \rightarrow \Lambda,\Sigma^0,\Sigma^-$ transitions 
is the one of Ref.~\cite{Singh:2006xp}, with the exception of an additional factor $M_Y/(M+M_Y)$ in the definitions of $f_2(q^2)$ and $g_2(q^2)$. In the limit of unbroken SU(3) symmetry, they can be determined from the experimental data on semileptonic decays of nucleons and hyperons. Assuming G-invariance, SU(3) symmetry and the conservation of vector current lead to $f_3(q^2)=g_2(q^2)=0$. We will also neglect $g_3(q^2)$, as its contribution is suppressed by mass of the outgoing lepton. This enables us to express the transition form factors in terms of $G_E^{p,n}$ and $G_M^{p,n}$,
the proton and neutron electric and magnetic form factors, respectively. Following Ref.~\cite{Singh:2006xp},  we introduce the vector and axial $N\rightarrow N$ form factors
\be
f_1^{p,n}(q^2) &=& \frac{G_E^{p,n}(q^2)+\tau G_M^{p,n}(q^2)}{1+\tau}\nonumber \\
f_2^{p,n}(q^2) &=& \frac{G_M^{p,n}(q^2)- G_E^{p,n}(q^2)}{1+\tau} \nonumber \\
g_A(q^2) &=& \frac{g_A}{(1-q^2/M_A^2)^2}
\label{N:form:fact}
\ee
where $\tau =-{q^2}/(4M)$, $M_A=1.05$ GeV, and $g_A=1.257$. As for $G_E^{p,n}$ and $G_M^{p,n}$ we employ the dipole parametrization
\begin{align}
G_E^p(q^2)&=\frac{1}{(1-q^2/M_V^2)^2} \nonumber\\
G_M^p(q^2)&=(1+\mu_p) G_E^p(q^2)\nonumber \\
G_E^n(q^2)&=-\frac{\mu_n \tau}{1+\lambda_n\tau} G_E^p(q^2) \nonumber\\
G_M^n(q^2)& = \mu_n G_E^p(q^2)\, ,
\end{align}
with $\mu_p= 1.7928$, $\mu_n=-1.9113$, $M_V=0.84$ GeV and $\lambda_n=5.6$. Using SU(3) symmetry  the form factors for the $N\rightarrow Y$ transitions are obtained from those introduced in Eq.~\eqref{N:form:fact}, and are listed in Table \ref{table:ff}.

\subsubsection{$N \rightarrow \Lambda_c$ transition}\label{sec:lambdaC}
 The study of charmed baryon production in weak processes is still in its infancy;
it involves nontrivial theoretical calculations and scarce experimental data are available.
The CHORUS collaboration recently measured the ratio of the cross section for $\Lambda_c$ production
in neutrino-nucleon charged-current (CC) interaction to the total charged-current cross section,
$\sigma(\Lambda_c)/\sigma(CC)=(1.54 \pm 0.35(\text{stat}) \pm 0.18(\text{syst})) \times 10^{-2}$ for a highly energetic
neutrino beam, $E_\nu=27$~GeV~\cite{KayisTopaksu:2003mh}.
However, in this energy region no reliable theoretical predictions for the form factors are currently available. 

Previous analyses on $\Lambda_c$ weak production in QE processes~\cite{Lellis:2004yn} do not seem to provide robust predictions for the total nuclear cross section. In fact, they rely on theoretical models for the $N\rightarrow \Lambda_c$ form factors whose predictions, never confronted with experimental data, differ by an order of magnitude for $E_\nu=10$ GeV. In this work, we compute the total cross section of $\Lambda_c$ weak production adopting form factors computed within four independent approaches: the CQMs of Refs.~\cite{Hussain:2017lir},  \cite{Gutsche:2014zna,Gutsche:2015rrt},  and \cite{PerezMarcial:1989yh}, and LQCD simulation of Refs.~\cite{Meinel:2016dqj,Meinel:2017ggx}. Their $q^2$ parameterizations, validated against the experimental data for the $\Lambda_c$ semileptonic decay reported by the  BESIII Collaboration~\cite{Ablikim:2015prg, Ablikim:2016vqd}, can be extended to the $q^2<0$ region, relevant for neutrino scattering. The same strategy is not applicable to the form factors of Ref.~\cite{Faustov:2016yza}, as they have poles for $q^2<0$.
As mentioned, the measurement of the $\Lambda_c\rightarrow \Lambda l^+\nu_l$ ($l=e,\mu$) decay width provides a constraint for the $\bar\nu_l\,\Lambda_c\rightarrow \Lambda\, l^+$  transition form factors. In the limit of unbroken SU(3) symmetry, the latter can be related to the $\bar\nu_l\,\Lambda_c\rightarrow n\, l^+$ form factors by performing a rotation in the SU(3) space -- the resulting Clebsh-Gordan coefficient being $\sqrt{3/2}$~\cite{Gutsche:2014zna}. In the following, we briefly review the various approaches employed to compute the form factors, providing the explicit expressions.

The CQM form factors obtained within the nonrelativistic harmonic-oscillator basis (NRHOQM) approach of Ref.~\cite{Hussain:2017lir} are calculated with the parameters for the baryon wave functions derived in Ref.~\cite{Roberts:2007ni}. They are used to compute the $\Lambda_c \to \Lambda$ semileptonic decay width, but not the  $\Lambda_c \to n $ transition. The calculated branching fractions are $\Gamma(\Lambda_c\to \Lambda l^+ \nu_l) / \Gamma_{\Lambda_c} = 3.84\%$ and 3.72\% for the electron and muon modes\footnote{$\Gamma_{\Lambda_c}$ is the total decay width of the $\Lambda_c$ hyperon.}, respectively. Though theoretical uncertainties were not provided, these predictions agree rather well with the experimental fractions,  $(3.63 \pm 0.38 \pm  0.20)\%$ [$e$] and $(3.49 \pm 0.46 \pm 0.27$) [$\mu$] measured by BESIII.  In addition, in Ref.~\cite{Hussain:2017lir} semileptonic decays to excited $\Lambda^*$ resonances  are calculated and leading order heavy-quark effective theory predictions are also  derived,  the latter being largely consistent with the quark-model form factors used in that work. Assuming SU(3) invariance, the form factors of the $\Lambda_c \to \Lambda$ transition can be used to estimate the weak production of the $\Lambda_c$ charm hyperon off a neutron by multiplying them by the appropriate Clebsh-Gordan coefficient. The matrix elements of the vector and axial $c\to s$ currents are expressed in Ref.~\cite{Hussain:2017lir} in terms of the $F_{1,2,3}(q^2)$ and $G_{1,2,3}(q^2)$ form factors, which are related to those introduced in Eq.~\eqref{eq:defffs} by 
\begin{align}
f_1(q^2)&=\sqrt{\frac{3}{2}}\bigg(F_1(q^2)+\nonumber\\[0.5em] 
&\left. \frac{M_{\Lambda_c}+M_\Lambda}{2M_{\Lambda_c}M_{\Lambda}} \left( M_{\Lambda}F_2(q^2)+ M_{\Lambda_c}F_3(q^2)\right)\right), \nonumber\\[0.5em] 
f_2(q^2)&=-\sqrt{\frac{3}{2}}\times\frac12\bigg(F_2(q^2)+ \frac{M_{\Lambda_c}}{M_\Lambda} F_3(q^2)  \bigg),\nonumber\\[0.5em]
g_1(q^2)&= \sqrt{\frac{3}{2}}\bigg(G_1(q^2)- \nonumber\\
&\left.\frac{M_{\Lambda_c}-M_\Lambda}{2M_{\Lambda_c}M_{\Lambda}} \left( M_{\Lambda}G_2(q^2)+ M_{\Lambda_c}G_3(q^2)\right)\right),\nonumber\\[0.5em]
g_2(q^2)&=\sqrt{\frac{3}{2}}\times\frac12\bigg( G_2(q^2)+ \frac{M_{\Lambda_c}}{M_\Lambda} G_3(q^2)  \bigg)
\label{eq:roberts_transf}
\end{align}
where the $\sqrt{3/2}$ Clebsh-Gordan coefficient has been explicitly included to use these for the $n\to\Lambda_c$ transition. The NRHOQM form factors of Ref.~\cite{Hussain:2017lir} have the simple form
\be
&\left.\begin{array}{ll}
F_{1,2,3}(q^2)\\[0.5em]
G_{1,2,3}(q^2)
\end{array}
\right\}  &= (A+Bq^2+Cq^4) e^{-\frac32\left(\frac{m_q p_\Lambda}{\alpha M_{\Lambda}}\right)^2}
\label{eq:roberts}
\ee
where $m_q=0.2848$ GeV,  $\alpha^2=(0.424^2+0.387^2)/2$ GeV and $p_\Lambda = \lambda^{1/2} (M_{\Lambda_c}^2, M_\Lambda^2, q^2)/2 M_{\Lambda_c}$, with $\lambda(x,y,z)\equiv x^2+y^2+z^2 -2xy - 2xz - 2yz$. The coefficients $A$, $B$, and $C$ can be found in Table~\ref{table:roberts}.

\begin{table}[h]
\begin{center}
\begin{tabularx}{0.4\textwidth}{ C{0.8} |C{0.8} |C{0.8} | C{0.8} }
\hline
\hline 
 & A & B [GeV$^{-2}$]& C [GeV$^{-4}$] \\[0.3em]
  \hline
 $F_1(q^2)$  & 1.382 &$-0.073$ &0  \\[0.3em]
 \hline
 $F_2(q^2)$  & $-0.235$ &0.022 &0.006  \\[0.3em]
 \hline
 $F_3(q^2)$  & $-0.146$ &$-0.003$ &$-0.001$  \\[0.3em]
 \hline
 $G_1(q^2)$  & 0.868 &0.013 &0.004  \\[0.3em]
 \hline
 $G_2(q^2)$  & $-0.440$ &$0.116$ &0.003  \\[0.3em]
 \hline
 $G_3(q^2)$  & 0.203 &-0.009 &0 \\[0.3em]
\hline
\end{tabularx}
\end{center}
\caption{Parameters introduced in Eq.~(\ref{eq:roberts}) to describe the $q^2-$dependence of the form factors used in the NRHOQM of Ref.~\cite{Hussain:2017lir}. Note that the we have changed the sign of the $B$ coefficient of the $G_2$ form factor, with respect to that quoted in Table III  of Ref.~\cite{Hussain:2017lir}. Only in this way, we could reproduce the form factor displayed in Fig.~4(a) of this latter reference. 
}
\label{table:roberts}
\end{table}

A relativistic CQM (RCQM) with infrared confinement~\cite{Ivanov:1996fj, Branz:2009cd, Gutsche:2013pp} is employed in Ref.~\cite{Gutsche:2014zna} to perform a detailed analysis of the $\Lambda_c\rightarrow n l^+\nu_l$ invariant and helicity amplitudes, form factors, angular decay distributions, decay width, and asymmetry parameters. The same scheme was later used~\cite{Gutsche:2015rrt} to predict the absolute branching fractions of $\Lambda_c\to \Lambda l^+ \nu_l$, which turned out to be 2.78\% and 2.69\% for the electron and muon channels, respectively -- no theoretical uncertainties are provided. These values are consistent with the lower limits of the data from the BESIII Collaboration~\cite{Ablikim:2015prg, Ablikim:2016vqd}. They also agree rather well with those ($2.9\pm 0.5 \%$ and $2.7\pm 0.6 \%$) quoted in the 2014 edition of the 
Review of Particle Physics~\cite{Agashe:2014kda}, obtained from the first model-independent measurement of the branching fraction of the $\Lambda_c^+\to p K^-\pi^+ $ mode, reported in 2013 by the Belle Collaboration~\cite{Zupanc:2013iki}. Within the RCQM, the $q^2$ behavior of the form factors is well represented by a double-pole parametrization of the form
\be
&\left.\begin{array}{ll}
f_{1,2}(q^2)\\[0.5em]
g_{1,2}(q^2)
\end{array}
\right\}  &=\frac{A}{1-B(q^2/M_{\Lambda_c}^2)+C(q^2/M_{\Lambda_c}^2)^2}\,.
\label{eq:gutsche}
\ee
For convenience, the parameters $A$, $B$, and $C$, taken form these references, are reported in Table~\ref{table:gutsche}. 

\begin{table}[h]
\begin{center}
\begin{tabularx}{0.4\textwidth}{ C{0.8} |C{1} |C{0.8} | C{0.8} }
\hline
\hline 
&   A & B & C   \\[0.3em]
\hline
  $f_1(q^2)$  &  0.470& 1.111&  0.303 \\[0.3em]
&0.511& 1.736 &0.760 \\[0.3em]
\hline
  $f_2(q^2)$ & $0.247$ & 1.240 & 0.390 \\[0.3em]
 &  $0.289$ & 1.970 & 1.054\\[0.3em]
\hline
  $g_1(q^2)$  & 0.414&  0.978 & 0.235 \\[0.3em]
 &  0.466 & 1.594& 0.647 \\[0.3em]
\hline
  $g_2(q^2)$ &  $-$0.073 & 0.781 & 0.225 \\[0.3em]
&  0.025 & 0.321 & 8.127  \\[0.3em]
\hline
\end{tabularx}
\end{center}
\caption{Coefficients employed in Eq.~(\ref{eq:gutsche}) to construct the   weak $\Lambda_c$ production form factors from the RCQM results of Refs.~\cite{Gutsche:2014zna, Gutsche:2015rrt}. For each form factor, the upper (lower) value corresponds to the $\Lambda_c\rightarrow N$ ($\Lambda_c\rightarrow \Lambda$) transition. Note that $g_2(q^2)=-f_2^A(q^2)$, being the latter form factor calculated in  \cite{Gutsche:2014zna, Gutsche:2015rrt}.}
\label{table:gutsche}
\end{table}

The results reported in Ref.~\cite{PerezMarcial:1989yh} were also obtained within a quark-model picture, improving previous predictions~\cite{AvilaAoki:1989yi} for the semileptonic decays of $\frac12^+$charm baryons, belonging to the representation {\bf 20} of SU(4), obtained in the four-flavor symmetric limit. Some SU(4) symmetry-breaking corrections, calculated within the MIT bag model (MBM)~\cite{Chodos:1974je, Chodos:1974pn} and a nonrelativistic quark model (NRQM)~\cite{Kokkedee}, are included. The main difference between MBM and NRQM lies in how the initial and final baryon overlap integrations are accounted for: they are limited to the bag space in MBM while in the NRQM they are not.  The MBM is expected to be more realistic than the NRQM at the price of introducing additional parameters, such as the bag radius and the masses of the quarks in the bag.  Additional corrections to the NRQM and MBM predictions, such as the hard-gluon QCD contributions~\cite{PerezMarcial:1989yh}, are also encompassed. Both the monopole and the dipole $q^2$ dependencies of the form factors have been employed within the MBM and the NRQM. For simplicity we only consider the latter, since monopole form factors overestimate the experimental $\Lambda_c\rightarrow \Lambda e^+\nu_e$ branching ratio. Dipole form factors provide reasonable estimates for this observable: 3.2\% and 3.9\% for the MBM and the NRQM, respectively.  They have the general form
\be
&\left.\begin{array}{ll}
f_{1,2}(q^2)\\[0.5em]
g_{1,2}(q^2)
\end{array}
\right\}  & = \frac{A}{(1-q^2/M_R^2)^2},
\label{eq:avila}
\ee
where the values of $A$ and $M_R$ are listed in Table~\ref{table:avila}.

\begin{table}[h]
\begin{center}
\begin{tabularx}{0.4\textwidth}{ C{1} |C{1} |C{1} | C{1}| C{1} }
\hline
\hline 
&NRQM  & & MBM & \\[0.3em]
\hline
   & $A$ & $M_R$ [GeV] & $A$ &  $M_R$ [GeV]  \\[0.3em]
  \hline
  $f_1(q^2)$  & 0.22 & 2.01 & 0.33 & 2.01 \\[0.3em]
 &0.35  & 2.11 & 0.46  & 2.11 \\[0.3em]
 \hline
   $f_2(q^2)$ & $0.11$& 2.01 &$0.18$ & 2.01  \\[0.3em]
 & $0.09$ & 2.11 & $0.19$ & 2.11  \\[0.3em]
 \hline
  $g_1(q^2)$ & 0.58  & 2.42 & 0.41 & 2.42 \\[0.3em]
& 0.61 & 2.51 & 0.50 & 2.51 \\[0.3em]
\hline
  $g_2(q^2)$ & 0.04  & 2.42 & 0.07 & 2.42  \\[0.3em]
 & 0.04 & 2.51 & 0.05 & 2.51 \\[0.3em]
\hline
\end{tabularx}
\end{center}
\caption{ Coefficients employed in Eq.~(\ref{eq:avila}) to construct the  weak $\Lambda_c$ production form factors 
for the MBM and NRQM models of Refs.~\cite{PerezMarcial:1989yh, AvilaAoki:1989yi}. For each form factor, the upper (lower) value corresponds to the $\Lambda_c\rightarrow N$ ($\Lambda_c\rightarrow \Lambda$) transition. The values of the form factors at $q^2=0$ are taken from ~\cite{PerezMarcial:1989yh}, and contain some SU(4) breaking corrections, while the pole masses collected in the table are taken from \cite{AvilaAoki:1989yi}. Note that $g_2(q^2)=-g_2^A(q^2)$, being the latter form factors calculated in  \cite{PerezMarcial:1989yh}.}
\label{table:avila}
\end{table}

Recently, the first LQCD results for both $\Lambda_c\rightarrow N$ and $\Lambda_c\rightarrow \Lambda$ form factors have become available \cite{Meinel:2016dqj,Meinel:2017ggx}. 
These works are based on gauge field configurations
generated by the RBC and UKQCD collaborations~\cite{Aoki:2010dy, Blum:2014tka} with 2+1 flavors of dynamical domain-wall fermions, for lattice spacings $a\approx 0.11$ fm and 0.085 fm and pion masses in the range $230 \,{\rm MeV}\, \le m_\pi \le 350\,{\rm MeV}$, although in  Ref.~\cite{Meinel:2016dqj} an additional ensemble  with $m_\pi = 139(2)$ MeV was also considered. The form factors are extrapolated to the continuum limit and to the physical pion mass employing a modified $z-$expansions~\cite{Bourrely:2008za}. Taking the Cabibbo-Kobayashi-Maskawa matrix element $|V_{cs}|$ from a unitarity global fit and the $\Lambda_c$ lifetime from experiments, branching fractions  ${\cal B}(\Lambda_c\rightarrow \Lambda e^+\nu_e)= \left(3.80\pm 0.19\pm 0.11\right)\%$ and ${\cal B}(\Lambda_c\rightarrow \Lambda \mu^+\nu_\mu)= \left(3.69\pm 0.19 \pm 0.11\right) \% $ were found in ~\cite{Meinel:2016dqj},  consistent with,  and twice more precise than, the BESIII measurements. The LQCD weak transition matrix elements are parameterized in terms of the $f_{\perp,\,+}(q^2)$ and $g_{\perp,\,+}(q^2)$ form factors, which are related to those defined in Eqs.~(\ref{jmu:1n}) and (\ref{eq:defffs}) for the $\Lambda_c$ weak production  in the following way\footnote{The additional form factors, $f_0$ and $g_0$, computed within LQCD, which only contribute to $f_3$ and $g_3$ in Eq.~(\ref{eq:defffs}), are not reported, as they do not contribute to the cross section in the limit of vanishing lepton mass.}:
\be
&f_1(q^2) &= -\frac{q^2}{s_+(q^2)}f_{\perp}(q^2)+\frac{(M_{\Lambda_c}+M_1)^2 } {s_+(q^2)}f_+(q^2) ,\nonumber \\[0.5em]
&f_2(q^2) &=  \frac{ M_{\Lambda_c}(M_{\Lambda_c}+M_1)}{s_+(q^2)} \bigg(f_{\perp}(q^2)-f_+(q^2)\bigg), \nonumber \\[0.5em]
&g_1(q^2) &=  -\frac{q^2}{s_-(q^2)}g_{\perp}(q^2)+\frac{(M_{\Lambda_c}-M_1)^2 } {s_-(q^2)}g_+(q^2), \nonumber \\[0.5em]
&g_2(q^2) &= \frac{ M_{\Lambda_c}(M_{\Lambda_c}-M_1)}{s_-(q^2)} \bigg(g_{\perp}(q^2)-g_+(q^2)\bigg)
\label{eq:lqcd_ff}
\ee
where $s_{\pm}(q^2) = \left[(M_{\Lambda_c}\pm M_1)^2-q^2\right]$, and $M_1$ is either $M$ or $M_\Lambda$, depending on the $\Lambda_c$ weak transition considered. The $q^2$ behavior is parameterized as
\be
&\left.\begin{array}{ll}
f_{\perp,+}(q^2)\\[0.5em]
g_{\perp,+}(q^2)
\end{array}
\right\}
&=\frac{1}{1-q^2/M_R^2}\big( A+B z(q^2)+C z^2(q^2) \big), \nonumber \\[0.5em]
&z(q^2) &= \frac{\sqrt{t_+-q^2} - \sqrt{t_+-t_0}}{\sqrt{t_+-q^2} + \sqrt{t_+-t_0}}, \nonumber \\[0.5em]
&t_+&=\left\{
\begin{array}{ll}
(M_D+M_K)^2\ & {\rm \ for }\  c\to s \\[0.5em]
(M_D+M_\pi)^2  & {\rm \ for}\  c\to d 
\end{array}
\right.
\label{eq:lqcd_ff_param}
\ee
with $t_0=(M_{\Lambda_c}-M_1)^2$, $ M_D=1.87\ {\rm GeV}, M_K=0.494\ {\rm GeV}$ and $M_\pi=0.135\ {\rm GeV}$ and the $A$, $B$, $C$ and $M_R$ coefficients 
are taken from the nominal fits carried out in Refs.~\cite{Meinel:2016dqj,Meinel:2017ggx}, providing the central values and statistical uncertainties of the form factors. 
We list all these parameters in Table~\ref{table:lqcd}.

\begin{table}[h]
\begin{center}
\begin{tabularx}{0.5\textwidth}{ C{0.8} |C{1.1} |C{1.1} | C{1.1}| C{0.9} }
\hline
\hline
   & A & B & C &  $M_R$ [GeV]  \\[0.3em]
  \hline
    $f_\perp(q^2)$  & $1.36\pm0.07$ & $-1.70\pm0.83$ & $0.71\pm4.34$ &  2.01 \\[0.3em]
   & $1.30\pm0.06$ & $-3.27\pm1.18$ & $7.16\pm11.6$ & 2.112 \\[0.3em]
 
 \hline
   $f_+(q^2)$  & $0.83\pm0.04$ & $-2.33\pm0.56$ & $8.41\pm3.05$ & 2.01 \\[0.3em]
 & $0.81\pm0.03$ & $-2.89\pm0.52$ & $7.82\pm4.53$ & 2.112 \\[0.3em]
 \hline
  $g_\perp(q^2)$ & $0.69\pm0.02$ & $-0.68\pm0.32$ & $0.70\pm2.18$ & 2.423 \\[0.3em]
  & $0.68\pm0.02$ & $-1.91\pm0.35$ & $6.24\pm4.89$ & 2.46 \\[0.3em]
\hline
  $g_+(q^2)$ & $0.69\pm0.02$ & $-0.90\pm0.29$ & $2.25\pm1.90$ & 2.423 \\[0.3em]
 & $0.68\pm0.02$ & $-2.44\pm0.25$ & $13.7\pm2.15$ & 2.46 \\[0.3em]
\hline
\end{tabularx}
\end{center}
\caption{Coefficients, taken from the nominal fits carried out in Refs.~\cite{Meinel:2016dqj,Meinel:2017ggx}, of the LQCD form factors used  in the parametrization of Eq.~(\ref{eq:lqcd_ff_param}). For each form factor, the upper (lower) value corresponds to the $\Lambda_c\rightarrow N$ ($\Lambda_c\rightarrow \Lambda$) transition.}
\label{table:lqcd}
\end{table}
 
\begin{figure*}[t]
\includegraphics[scale=0.8]{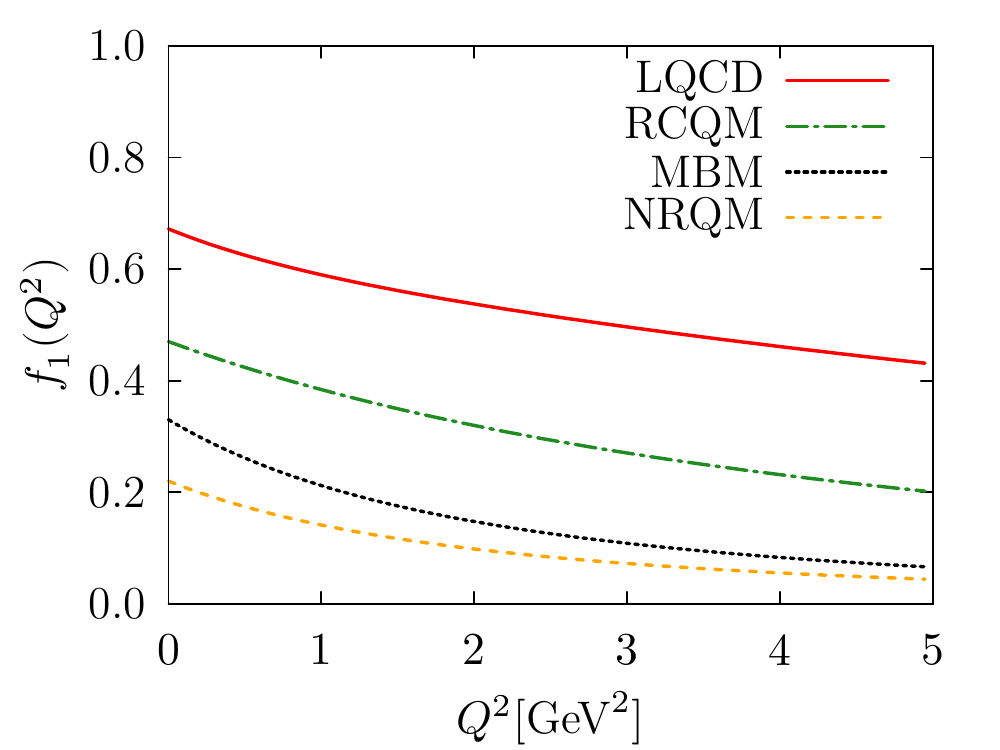}
\includegraphics[scale=0.8]{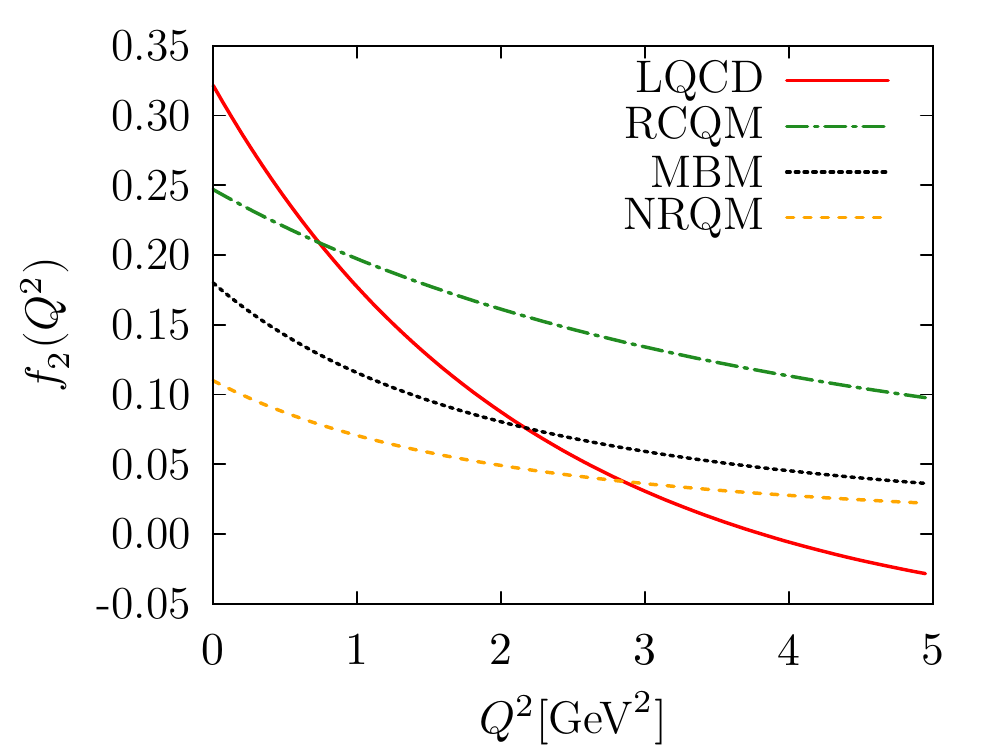}
\includegraphics[scale=0.8]{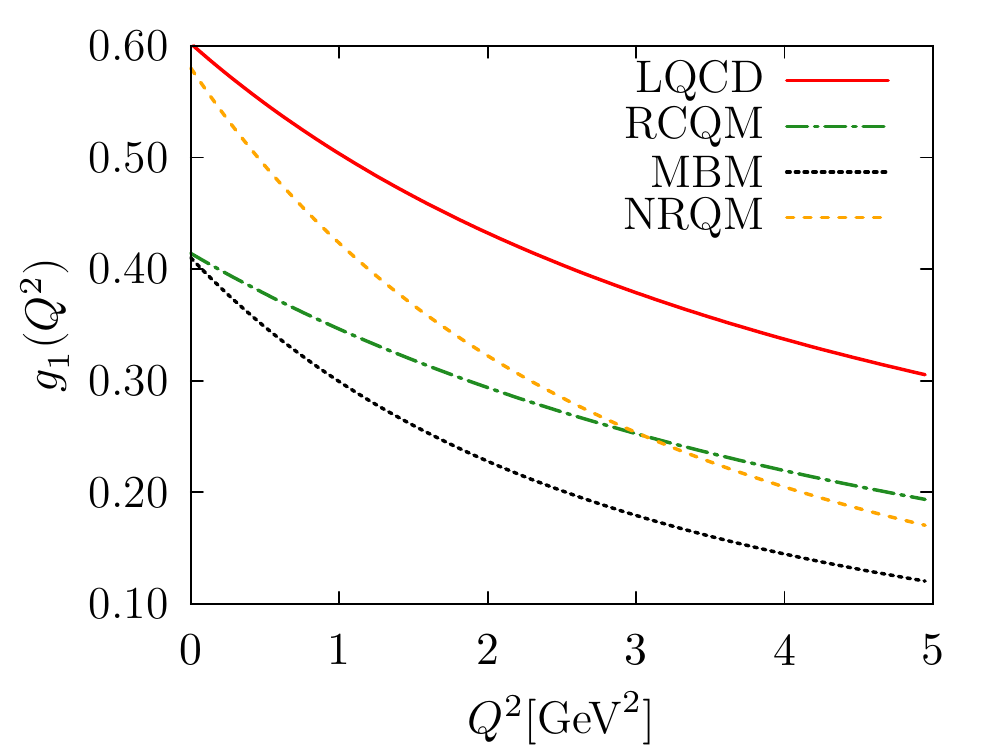}
\includegraphics[scale=0.8]{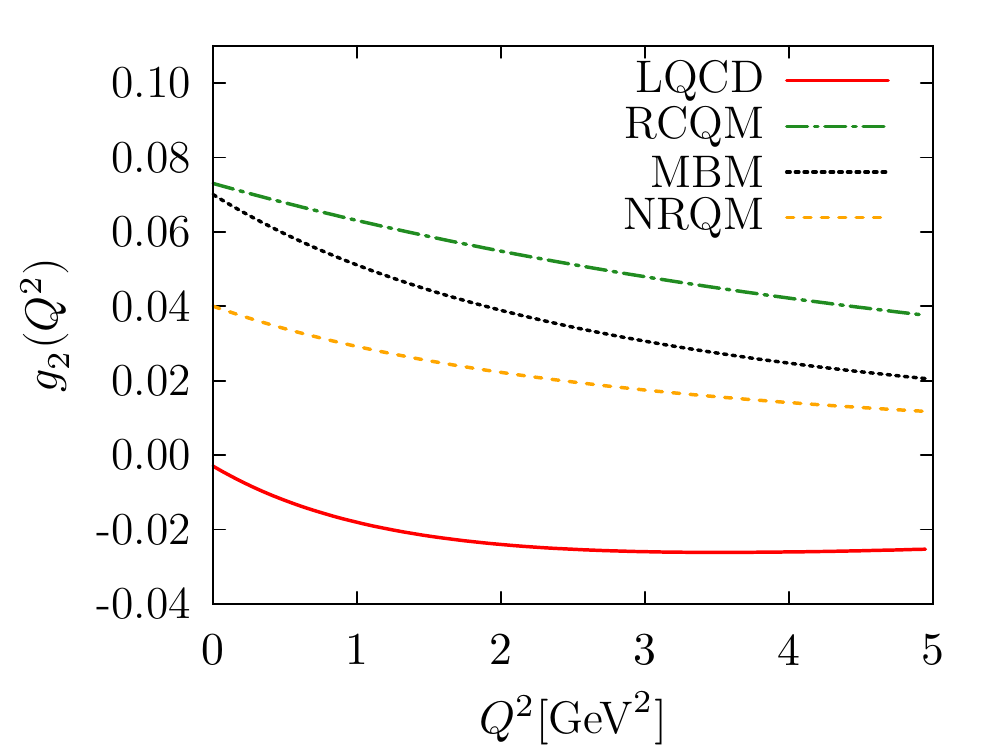}
\caption{Form factors for the $n\rightarrow \Lambda_c$ transition deduced from extrapolating to $Q^2=-q^2>0$ the results reported in 
Refs.~\cite{Meinel:2017ggx} (LQCD), \cite{Gutsche:2014zna} (RCQM) and  \cite{PerezMarcial:1989yh} (MBM and NRQM).}
 \label{fig:ff_charmNuc}
\end{figure*}
 
The semileptonic branching ratios $\mathcal{BR}( \Lambda_c \rightarrow \Lambda e^+\nu_e)$ calculated with the five different sets of form factors turn out to be in a reasonable agreement with each other and with the BESIII measurement. The predictions range from the 2.78\% of the RCQM~\cite{Gutsche:2014zna}, to the 3.9\% obtained within the NRQM scheme~\cite{PerezMarcial:1989yh}, corresponding to a $\simeq 30\%$ variation. Hence, one should expect a similar spread among the predictions for the total $\nu_l + n\rightarrow \Lambda_c^+ +l^-$ cross section. However, the theoretical uncertainty on the latter observables could be even larger for the following reasons. (i) Experimental measurements effectively constraint the $\Lambda_c\rightarrow \Lambda$ form factors, while we are interested in the $n \rightarrow \Lambda_c$ transition. The form factors for $\Lambda_c\rightarrow \Lambda$ and $n \rightarrow \Lambda_c $ transitions may be related by an SU(3) rotation but, because of SU(3) breaking effects, these are subject to additional corrections. In this regard, one has to note that for most of the models we have also access to  direct predictions for the $c\to d$ transition form factors. (ii) The form factors have been fitted to the experimental decay width in the range $0 \leq q^2 \leq (M_{\Lambda_c}-M_{\Lambda})^2$, while $\Lambda_c$ neutrino production is characterized by $q^2<0$. The extrapolation of the form factors to moderately large negative $q^2$ values, corresponding to medium-energy neutrino beams, entails a certain degree of ambiguity. In order to quantify the uncertainty of this procedure, we consider five different sets of form factors characterized by a variety of $q^2$ dependencies. (iii) The $\Lambda_c \rightarrow \Lambda e^+\nu_e$ decay is mostly sensitive to the axial $g_1(q^2)$ form factor, which dominates the total width. On the other hand, both $f_1(q^2)$ and $g_1(q^2)$ are important in the $\nu_l +n\rightarrow \Lambda_c ^+ + l^-$ process. Thus, the $f_1(q^2)$ form factor, being less constrained by the experiment, may introduce an additional error to the cross-section prediction.

To evaluate the $\nu_l + n\rightarrow \Lambda_c^+ +l^-$ cross section for the LQCD~\cite{Meinel:2016dqj,Meinel:2017ggx}, the RCQM of Refs.~\cite{Gutsche:2014zna, Gutsche:2015rrt} and the MBM \& NRQM~\cite{PerezMarcial:1989yh} schemes, we pursue two different avenues. The first one corresponds to the unbroken SU(3) limit, in which we take the $\Lambda_c\to \Lambda$ form factors and use the $\sqrt{3/2}$ Clebsh-Gordan coefficient to compute the $\Lambda_c$ production off the neutron. In the second one, we directly apply the $\Lambda_c\rightarrow N$ form factors reported in these references.  As for the NRHOQM of Ref.~\cite{Hussain:2017lir}, the only available parametrization, obtained from the SU(3) rotation, will be employed. 

In Fig.~\ref{fig:ff_charmNuc}, we show the form factors for the $N\rightarrow \Lambda_c$ transition, for $Q^2=-q^2>0$, corresponding to the kinematics of neutrino production. As we mentioned, the main contribution to the cross section and to the decay width is driven by $f_1(Q^2)$ and $g_1(Q^2)$. LQCD predicts the largest values for these two form factors, while the MBM and NRQM models provide the far lowest ones for $f_1(Q^2)$ and, at  high $Q^2$, for $g_1(Q^2)$. Note that the LQCD and RCQM form factors exhibit similar $Q^2$ dependencies, except for $f_2$,  being the $f_1$ and $g_1$ from RCQM form factors $\sim30\%$ smaller than those computed within LQCD. Analogous behaviors are observed for the $\Lambda_c\rightarrow \Lambda$ form factors, discussed in detail in the Appendix, where further comparisons among the various approaches are provided.

As a final remark, analogously to the case of $\Lambda$ and $\Sigma$ production, the $f_3$ and $g_3$ form factors have been neglected in the cross-section calculations, as they are suppressed by a factor $m_\ell^2/M^2$.
%

\section{Nuclear effects}
\label{sec:sfs}
In this Section we generalize the discussion of Sec.~\ref{nu:nucl:scatt} to the case in which the $\bar{\nu}$ beam scatters off a nucleus with $A$ nucleons producing a hadronic final state comprised of a hyperon $Y$ and an $(A-1)$-nucleon residual system. In the kinematical region in which the impulse approximation is expected to be applicable, the nuclear matrix element can be still expressed as in Eqs.~\eqref{jmu:1n} and \eqref{eq:defffs}, provided that the elementary interactions occur on single bound nucleons. In what follows, we will restrict the discussion to isospin-symmetric nuclei, for which we assume the neutron and proton densities to coincide. Within this scheme, the double-differential cross section is evaluated as in Eq.~(\ref{eq:sigma2D}), but the hadron tensor is obtained from the convolution of the hole SF and the spin averaged squared amplitude of the hadron matrix element, $A^{\mu\nu}$, reported in Eq.~\eqref{eq:amunu}
\begin{align}
W^{\mu\nu}(q) = &  \int \frac{d^3p}{(2\pi)^3} \int dE S_h(E,\vec{p}\,)  \frac{M}{E_p}\frac{M_Y}{E_{p+q}^Y} \nonumber\\
&\times \delta(E+q^0-E_{p+q}^Y)A^{\mu\nu}(p,q)\, .
\label{eq:hadron_tens_ia}
\end{align}
We note that the hole SF, $S_h(E, \vec{p}\,)$, yields the probability density of finding a nucleon in the target nucleus with a given momentum $\vec{p}$ and removal energy $E$.

\subsection{Nucleon spectral function}\label{sec:sf}
In Ref.~\cite{Singh:2006xp}, the local Fermi gas (LFG) model was adopted to describe nuclear dynamics. This amounts to rewriting the hole SF as
\beq 
S_h^{\rm LFG}(E,\vec{p}\,)=\int d^3r \frac{\rho(r)}{2} \frac{6\pi^2}{k_F^3 (r)} \theta(k_F(r)-|\vec{p}\,|)\delta(E-E_p)\,,
\label{LFG:sf}
\eeq
where $k_F(r)=(3\pi^2\rho(r)/2)^{1/3}$ is the Fermi momentum and $\rho(r)$ is the total point-nucleon density distribution (protons plus neutrons). 

It has long been known that models based on an independent-particle description of the nuclear structure largely fail to account for the complexity of nucleon correlations in nuclei. In this work, we improved the model of Ref.~\cite{Singh:2006xp} by considering two different realistic hole SFs: a semi-phenomenological one from the Valencia group (further referred to as LDA-SF) based on the findings of Ref.~\cite{FernandezdeCordoba:1991wf}, and a second one obtained within the Correlated Basis Function (CBF) theory. For a detailed description of both models we refer to Ref.~\cite{Sobczyk:2017vdy}. In that work, the differences between the two SFs were analyzed by comparing the electromagnetic scaling functions of $^{12}$C, and an overall good agreement was found in all the kinematical setups considered. 

In analogy to Eq.~\eqref{LFG:sf}, the LDA-SF of finite nuclei is obtained through the local density approximation (LDA) procedure. The infinite nuclear matter hole SF derived in \cite{FernandezdeCordoba:1991wf}, denoted by $S_{h,\text{NM}}^{\rm LDA}(E,\vec{p},\rho)$, is calculated for different values of the nuclear density and integrated over the density profile of the nucleus
\begin{align}
S_h^{\rm LDA}(E,\vec{p}\,)&=\int d^3r \frac{\rho(r)}{2} \frac{6\pi^2}{k_F^3(r)} S_{h,\text{NM}}^{\rm LDA}(E,\vec{p},\rho) \nonumber \\
&=2 \int d^3r S_{h,\text{NM}}^{\rm LDA}(E,\vec{p},\rho)\, . \label{LDA:sf}
\end{align}
However, when computing the cross section, the integration over the nuclear volume is carried out for the full hadron tensor~\cite{Nieves:2004wx,Nieves:2017lij}
\begin{align}
 W^{\mu\nu}_{LDA}(q) &=  2 \int d^3r  \int \frac{d^3p}{(2\pi)^3} \int  dE \, S_{h, \text{NM}}^{\rm LDA}(E,\vec{p},\rho) \nonumber \\
 &\times \frac{M}{E_p}\frac{M_Y}{E^Y_{p+q}} \delta(E+q^0-E^Y_{p+q}(\rho))A^{\mu\nu}(p,q)\,. \label{eq:hadron_tens_lda}
\end{align}
Note that the same procedure is followed within the LFG model adopted in Ref.~\cite{Singh:2006xp}, and associated to Eq.~(\ref{LFG:sf}). 

Within the CBF approach~\cite{Benhar:1989aw,Benhar:1994hw}, the hole SF of finite nuclei entering Eq.~(\ref{eq:hadron_tens_ia}) is given by the sum of two contributions
\beq 
S_h^{\rm CBF}(E,\vec{p}\,)=S_h^{\rm MF}(E,\vec{p}\,)+S_h^{\rm corr}(E,\vec{p}\,)\, .
\label{CBF:sf}
\eeq
The first term is associated to the low momentum and removal-energy region. It is derived within a modified mean field (MF) scheme in which correlations are included through quenched spectroscopic factors, extracted from $(e,e'p)$ scattering measurements. The LDA is only adopted to determine $S_h^{\rm corr}(E,\vec{p}\,)$, corresponding to the high-energy and momentum region of the SF, which is largely unaffected by finite-size and shell effects, as it essentially arises from short-range nuclear dynamics.

\subsection{Final state interaction for hyperons}\label{sec:fsi}
The hole SF accounts for the complexity of nuclear interactions pertaining to the nucleon in the initial nuclear target. However, in the kinematical region in which the interactions between the struck particle and the spectator system cannot be neglected, the impulse-approximation results have to be corrected. In more standard electron- and neutrino-nucleus scattering calculations, where the struck particle is either a proton or a neutron, this is achieved by means of the particle SF~\cite{Nieves:2017lij,Ankowski:2014yfa}. Within the CBF theory, the real part of a phenomenological optical potential~\cite{Cooper:1993nx} is used to modify the energy spectrum of the outgoing nucleon. In addition, a convolution scheme that uses as input the nuclear transparency and nucleon-nucleon scattering amplitudes is adopted to redistribute the strength from the QE peak to the higher-energy regions. On the other hand, within the Valencia model, the particle SF is consistently obtained with the hole SF, although relativistic corrections can be included in the former. 

In this work, the interactions between the outgoing hyperon and the $A-1$ nucleons are described simply employing a phenomenological MF potential, to modify the hyperon energy spectrum as
 $\tilde{E}^Y_{p+q}(\rho)=E^Y_{p+q}+V(\rho)$,  where $V(\rho)=-30 \rho/\rho_0\, \text{MeV}$~\cite{Bouyssy:1979tb},  and $\rho_0=0.16\, {\rm fm}^{-3}$, the nuclear saturation density.  Within the LDA-SF, the density-dependent hyperon spectrum is included by replacing the energy conservation delta function in Eq.~\eqref{eq:hadron_tens_lda} by  $\delta(E+q^0-\tilde{E}^Y_{p+q}(\rho))$. Nuclear modifications affecting the 
nucleon tensor $A^{\mu\nu}$ are expected to be much smaller and have not been considered. On the other hand, within the CBF approach, we simply modify the energy-conserving $\delta$ function of Eq.~\eqref{eq:hadron_tens_ia} as $ \delta(E+q^0-\bar{E}^Y_{p+q})$ with the average hyperon spectrum
\beq
\bar{E}^Y_{p+q} = E^Y_{p+q} + \frac{1}{A}\int d^3 r \rho(r) V(\rho)\ .
\eeq

The hyperon produced in the primary vertex travels through the nuclear environment, interacting with  nucleons and exchanging momentum and possibly producing a different hyperon. Although a quantum-mechanical description of the scattering processes would be more appropriate, here we use a Monte Carlo algorithm analogous to that presented in Ref.~\cite{Singh:2006xp}, using as input the available measurements of hyperon-nucleon scattering cross sections. This amounts to treating collisions in a classical fashion, as in most of the available neutrino event generators~\cite{Hayato:2002sd,Juszczak:2005zs,Andreopoulos:2009rq}. 

According to the total initial cross sections for $\Sigma^{0}, \Sigma^-$ and $\Lambda$ production, we select the type of hyperon produced in the primary interaction vertex. Then, for this particular hyperon, we compute the 
differential cross section $d\sigma/(d\Omega(\hat{k}\,') d E_{k'} d^3r)$ or $d\sigma/(d\Omega(\hat{k}\,') d E_{k'} d|\vec{p}\,|)$, for the LDA-SF or the CBF-SF cases, respectively\footnote{ As for the LDA-SF approach, although the initial nucleon momentum is already integrated out, some effects of the hole SF are retained, and they modify the distribution profiles of the outgoing-lepton kinematics. Conversely, the CBF-SF scheme enables to sample the magnitude of the momentum of the initial nucleon to be used in the cascade algorithm.}. These are the weights of the events that are taken as input in the MCC. 

According to the calculated profiles, we randomly generate the charged lepton energy $ E_{k'}\in [m_\ell,E_\nu]$ and scattering angle $\theta^\prime \in [0,\pi]$, together with the position ${\vec{r}_1}=(r_1\cos\theta_1\sin\phi_1,r_1\sin\theta_1\sin\phi_1,r_1\cos\phi_1)$, in which the hyperon has been produced.  (Here $r_1 \in [0,r_{\rm max}]$, with $r_{\rm max}$ sufficiently large to safely take $\rho_{p(n)}(r_{max})=0$.) Since the elementary cross section obtained using the CBF-SF does not depend upon $r$, we generate $\vec{r}_1$ according to the density profile of the nucleus. When using the LDA-SF in the calculation of the cross section, we perform the integration over the initial nucleon momentum beforehand, as in Ref.~\cite{Singh:2006xp}. In this case, we assume that the momentum of the hyperon produced at the interaction vertex is $\vec{p}_{Y_1} = \vec{q}+\vec{p}_{\text{gen}}$, where $\vec{p}_{\text{gen}}$ is a randomly generated three-vector below the local Fermi sea ($|\vec{p}_{\text{gen}}|\le k_F(r)$). On the other hand, the CBF-SF allows one to consistently generate the momentum modulus of the struck nucleon on an event-by-event basis and to include it in the definition of $\vec{p}_{Y_1}$, choosing randomly an angle between $\vec{p}$ and $\vec{q}$ using a $]0,\pi]$ flat distribution. In both approaches, we assume that this initial hyperon is on-shell, and its energy is given by $E_{Y_1}=\sqrt{M_{Y_1}^2+ \vec{p}_{Y_1}^{\,2}}$.

We simulate the hyperon propagation from the interaction vertex until it exits the nucleus by iterating the following steps~\cite{Singh:2006xp}:
\begin{enumerate}
\itemsep0em
\item Assuming that the hyperon kinetic energy is significantly larger than $V(\rho)$, we propagate it by a short distance $\vec{dl}=\vec{p}_{Y_1}/E_{Y_1} dt$, $dt$ being a small time interval, along its momentum direction. We then randomly select a nucleon from below the Fermi sea, $|\vec{p}_1|<k_F$, and compute the invariant energy $E_{\rm inv}$ of the $Y_1+N_1$ system. We evaluate the interaction probability per unit length for the various scattering processes permitted by charge conservation $Y_1+N_1\rightarrow Y_i+N_i$, where $Y_i=\Lambda,\Sigma^-,\Sigma^0,\Sigma^+$ and $N_i=p,n$. 
For a given channel $i$, this is given by ${\cal P}_{i}=\left[\rho_p\sigma_{[Y_1+p\rightarrow Y_i+N_i]}(E_{\rm inv})+\rho_n\sigma_{[Y_1+n\rightarrow Y_i+N_i]}(E_{\rm inv})\right]$, where the total cross sections $\sigma_{[Y N\rightarrow Y' N' ]}(E_{\rm inv})$, extracted from the available experimental data, are compiled in the Appendix of Ref.~\cite{Singh:2006xp}. The probability that the interaction has occurred is ${\cal P}=\sum_i{\cal P}_i\, dl$. Note that $dt$ has to be small enough so that ${\cal P} dl \ll 1$. A random number $x \in [0,1]$ is generated. If $x> \sum_i{\cal P}_i\, dl$ the interaction has taken place and we proceed to the next step; otherwise we skip it and go to step 3.

\item We select the interaction channel according to their respective probabilities (cross sections). We generate a random angle for the production of the outgoing $(N_i, Y_i)$ pair in the center-of-mass of $N_1+Y_1$ system. We boost back to the laboratory frame and we implement Pauli blocking by checking that the momentum of the final nucleon is larger than $k_F(r)$. If this condition is satisfied, we have a new hyperon (possibly of different type) propagating with a new direction and momentum. If not, the interaction did not occur and the hyperon properties remain unchanged.

\item The new position of the hyperon is  $\vec{r}_i =\vec{r}_1+ \vec{d l}$. If no interaction has occurred, in this point of the nucleus, the three-momentum and type of hyperon correspond to those of the initial hyperon $Y_1$. Otherwise, they correspond to those of the outgoing hyperon $Y_i$ after the collision.

\item If the hyperon's kinetic energy is smaller than 30 MeV, we stop the propagation and  assume that the hyperon has exited the nucleus without any further interaction. We recall here that the MC code does not include the effects of the real part of the hyperon optical potentials and that straight-line trajectories are always assumed. Quantum effects, neglected in the cascade approach, are expected to become especially important at low energies.

\end{enumerate}
\begin{figure}[!htb]
\includegraphics[scale=0.7]{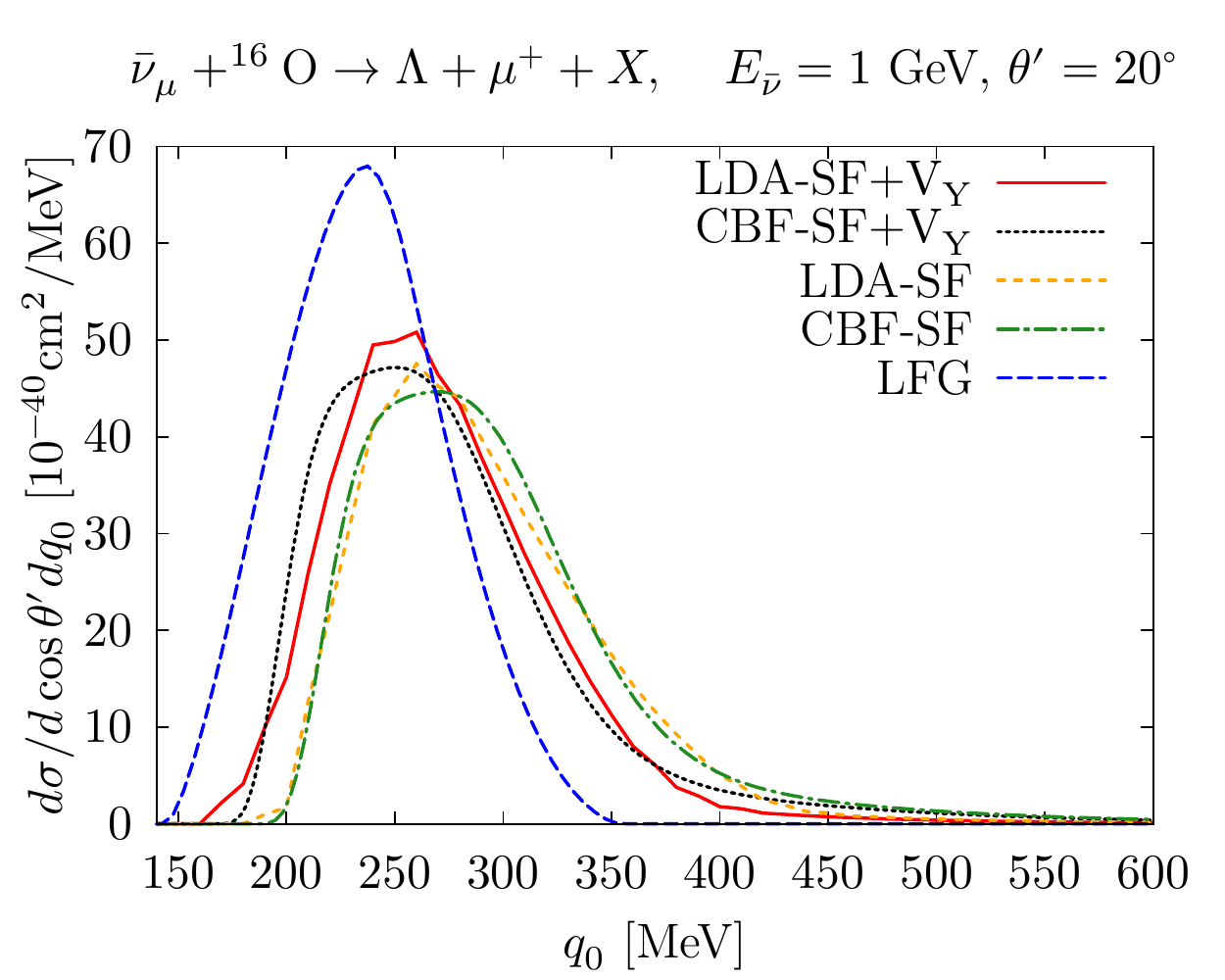}
\includegraphics[scale=0.7]{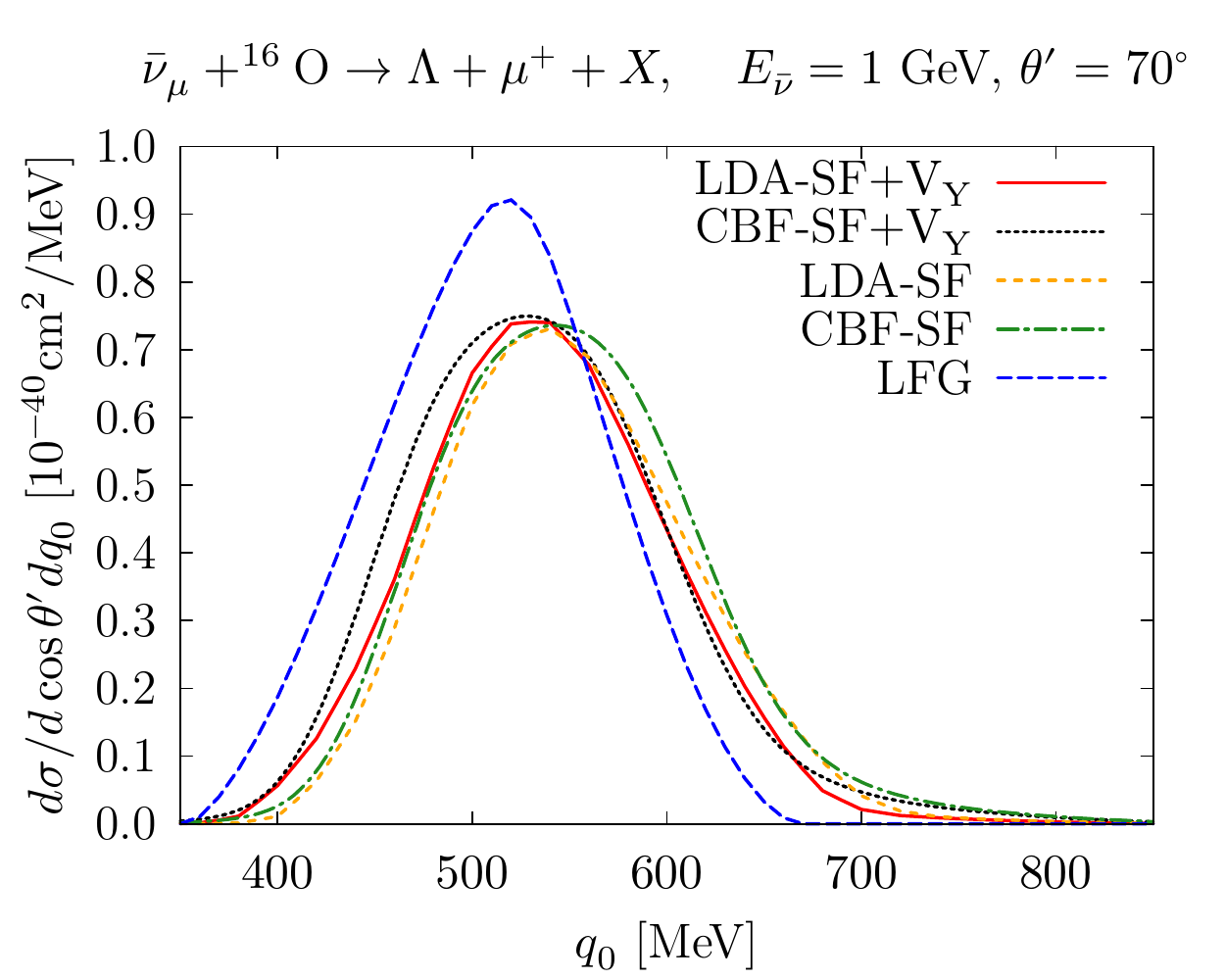}
\caption{Differential cross section $d^2\sigma(\bar{\nu}_{\mu}+^{16}{\rm O}\rightarrow \Lambda+\mu^++X)/(d\cos\theta' dq^0)$  for  $E_{\bar{\nu}}=1$ GeV and  
 two fixed antimuon scattering angles, $\theta'=20^\circ$ and  $\theta'=70^\circ$, showed in the top and bottom panels, respectively. The blue dashed, green dot-dashed and orange short dashed lines correspond to the LFG, CBF-SF and LDA-SF calculations, respectively. The red solid and black dotted curves stand for the LDA-SF and CBF-SF cross sections, when a mean-field potential is included in the energy spectrum of the hyperon. All the results have been obtained without employing the MCC. } \label{fig:dsigmadEdcosT}
\end{figure}
%
%
\section{Results}\label{sec:results}
\subsection{Strange hyperon production}

\begin{figure*}[tbh]
\includegraphics[scale=0.7]{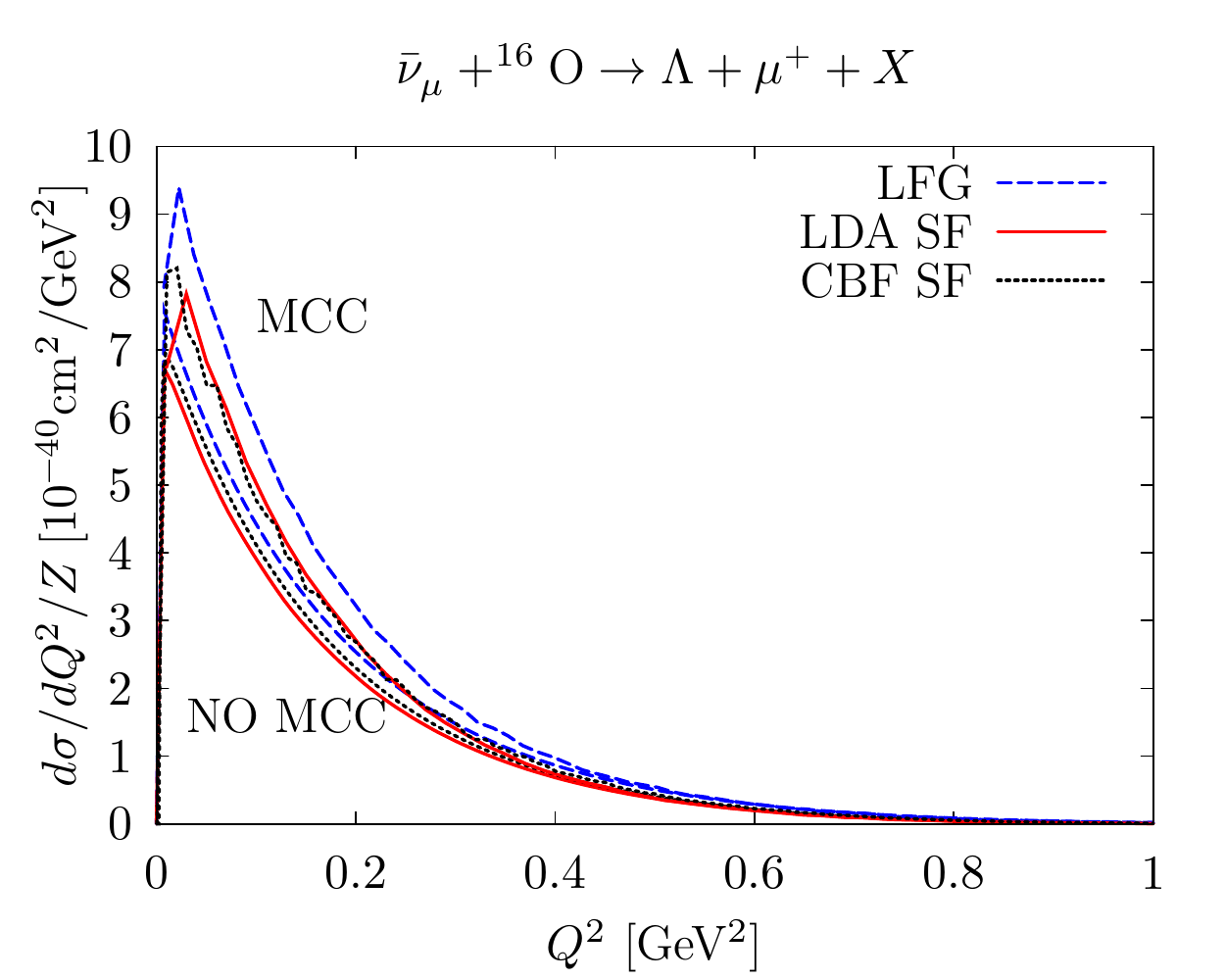}
\includegraphics[scale=0.7]{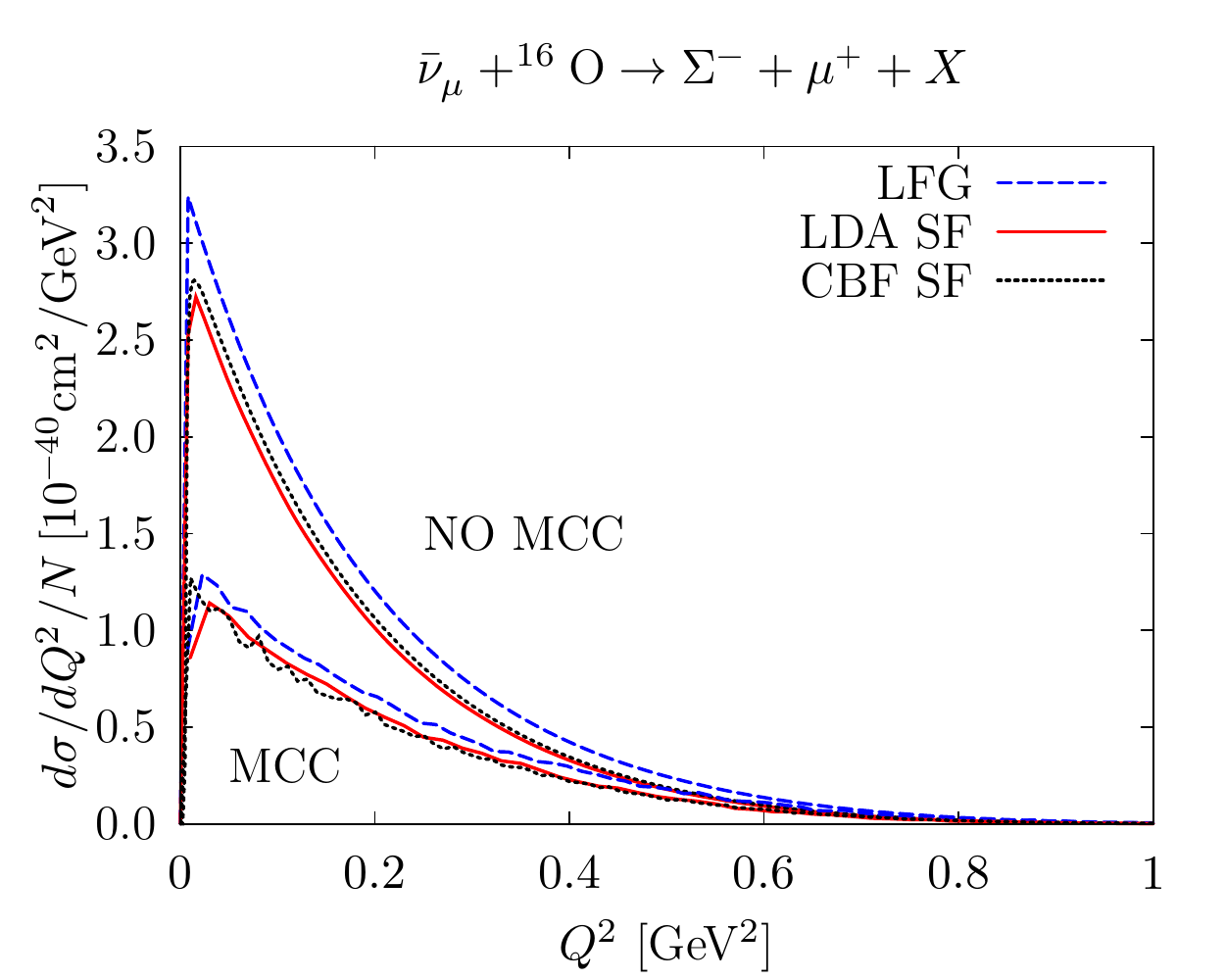}
\includegraphics[scale=0.7]{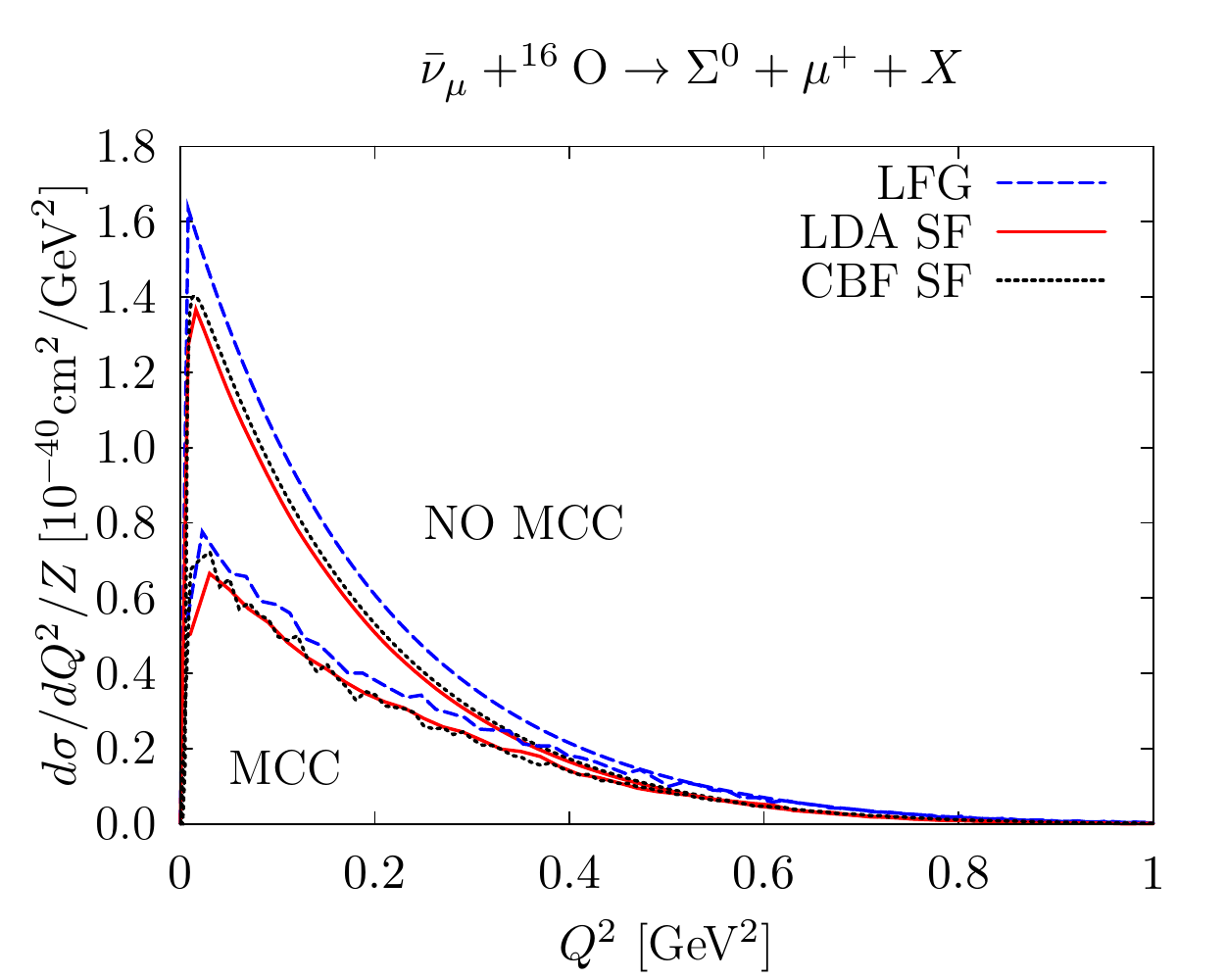}
\includegraphics[scale=0.7]{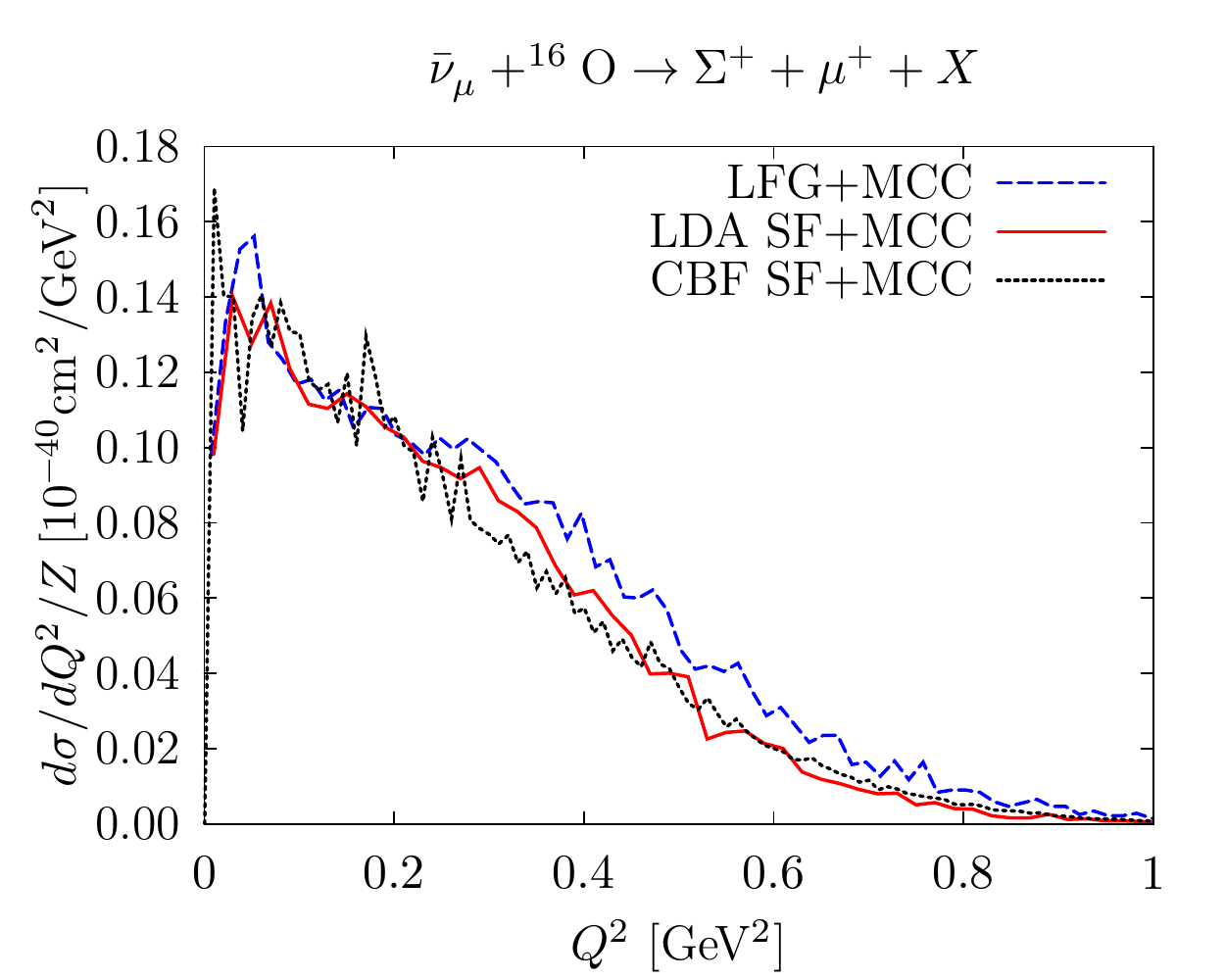}
\caption{$d\sigma /dQ^2$ cross sections, per number of target protons or neutrons, for $\Lambda$, $\Sigma^-$, $\Sigma^0$, and $\Sigma^+$ production in oxygen and  $E_{\bar{\nu}}=1$ GeV. 
For each type of hyperon, we show results from the LFG, CBF-SF and LDA-SF approaches, which are depicted as blue dashed, black dotted and red solid lines, respectively. The MCC and NO MCC labels denote whether the nuclear cascade algorithm has been applied or not.}
 \label{fig:dsigmadq2}
\end{figure*}

Let us first focus on the role played by the SFs in the description of the hyperons production. The double-differential cross section 
$d^2\sigma/(d\cos\theta' dq^0)$ of the process $\bar{\nu}_{\mu}+^{16}$O$\rightarrow \Lambda+\mu^++X$ is shown in Fig.~\ref{fig:dsigmadEdcosT} for different models of the hole SF, and two fixed outgoing lepton scattering angles. Note that the MCC has not been employed to obtain these results. We observe an overall good agreement between the  LDA-SF and the CBF-SF results. As expected, in both cases the hyperon mean-field potential produces a shift of $\sim 30$ MeV in the position of the QE peak, also leading to an enhancement of its height. The LFG calculations, represented by the blue dashed lines, have been carried out assuming a free energy spectrum for both the initial nucleon and the hyperon, as in Ref.~\cite{Singh:2006xp}.  The comparison with the other curves, in which a more realistic description of the nuclear dynamics is adopted, reveals that nuclear correlations sizably affect the inclusive cross sections. In particular, nuclear correlations in the initial state reduced the height of the QE peak, redistributing the strength to the higher energy-transfer. These effects are more apparent for $\theta'=20^\circ$ than for $\theta'=70^\circ$, being the cross section much bigger (between one to two orders of magnitude) for the lower scattering angle. The $q^0$ values relevant for the QE cross section increase with $\theta'$, and thus one should expect the largest cross sections for relatively small outgoing lepton scattering angles, where the effects of initial-state nuclear correlations are more important.

The differential cross sections, $d\sigma/dQ^2$, for $\Lambda, \Sigma^-,\Sigma^0$ and $\Sigma^+$ production from oxygen are shown in Fig.~\ref{fig:dsigmadq2}, for an incoming muon antineutrino energy of 1 GeV and the LFG, LDA-SF and CBF-SF approaches. We compare the results obtained either applying or not applying the MCC, the corresponding curve labelled as ``MCC'',  or ``NO MCC'', respectively. Nucleon-nucleon correlations, encoded in the realistic hole SFs, quench the cross sections in all the hyperon-production channels. However, their impact is far less dramatic than for the double-differential cross sections of Fig.~\ref{fig:dsigmadEdcosT}. Considerably more relevant are the effects of the MCC. They strongly modify the initial calculation leading to a non-zero $\Sigma^+$ cross section, to a sizable enhancement of the $\Lambda$ production and to a drastic reduction -- more than $50\%$ -- of the $\Sigma^0$ and $\Sigma^-$ distributions. This can be qualitatively understood by analyzing the kinematics of these processes. While traveling through the nucleus, the hyperons dissipate their kinetic energy in the scattering processes. The $\Sigma$ hyperons are heavier than the $\Lambda$ baryon, hence their production in the nuclear cascade is generally suppressed. In particular, for low energies the $\Lambda\rightarrow\Sigma$ process is kinematically forbidden. Therefore, for low-energy hyperons, the $\Sigma\rightarrow\Lambda$ processes dominate.  On the other hand, $\Sigma^+$ hyperons can be produced only in secondary collisions. 

\begin{figure*}[tbh]
\includegraphics[scale=0.7]{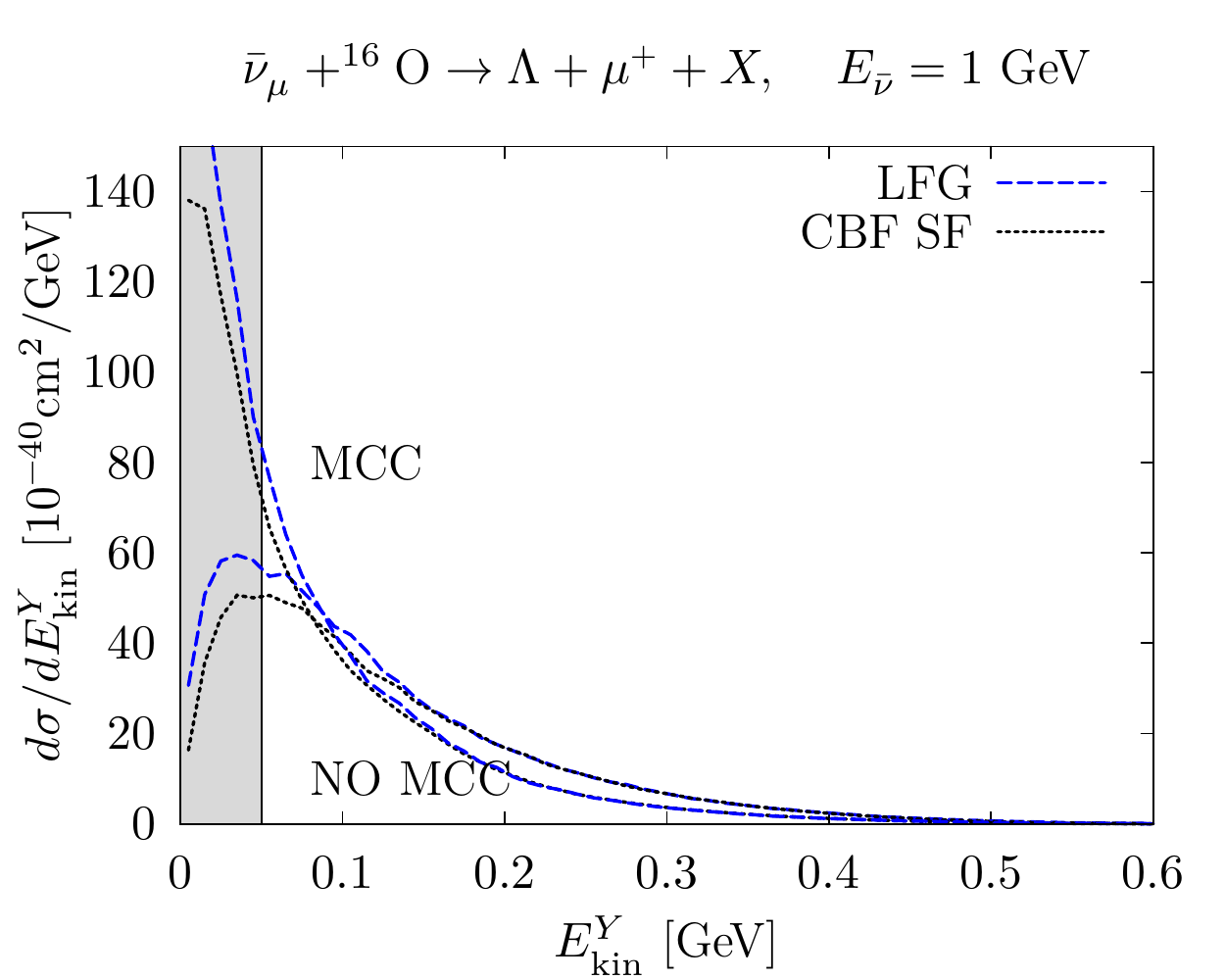}
\includegraphics[scale=0.7]{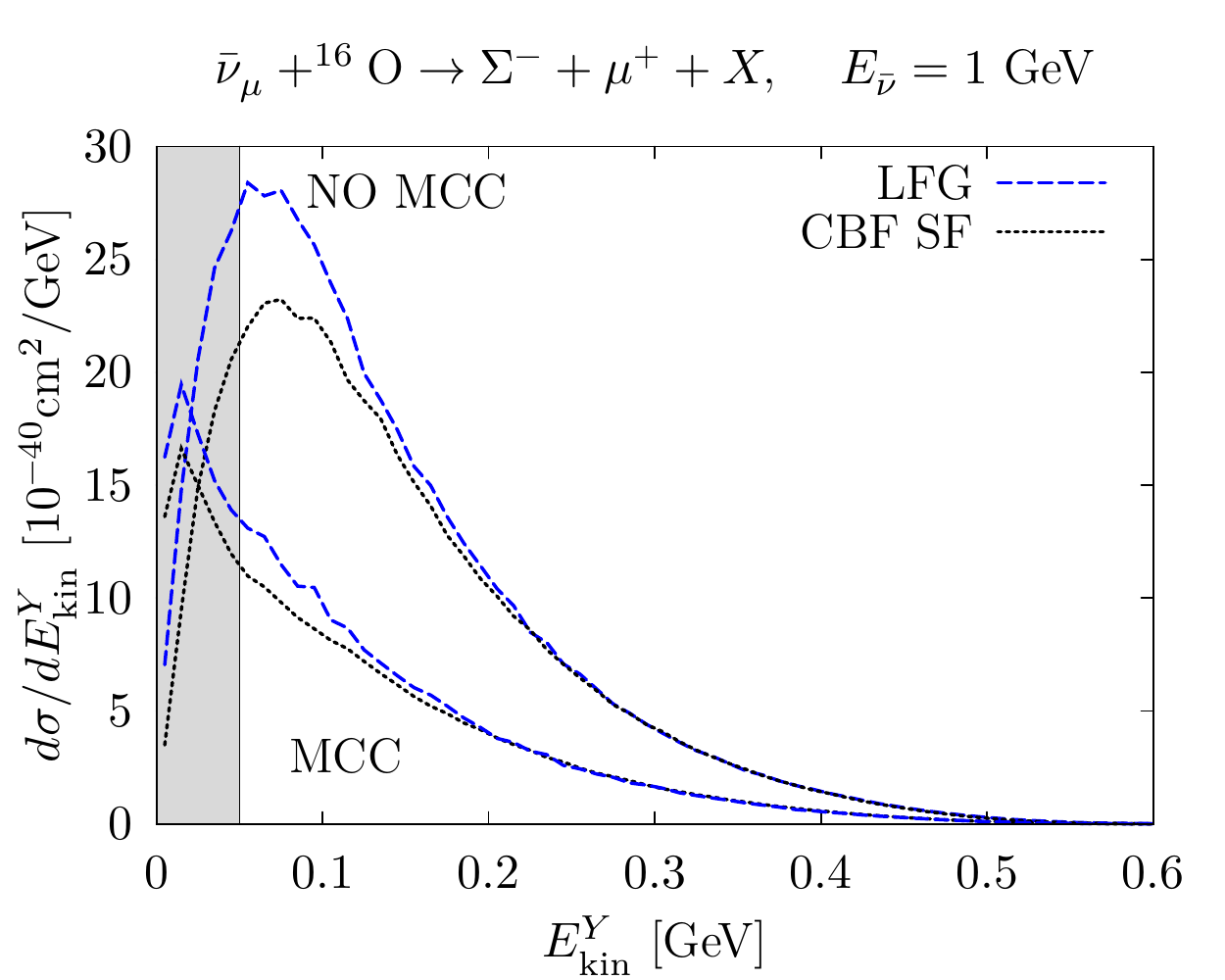}
\includegraphics[scale=0.7]{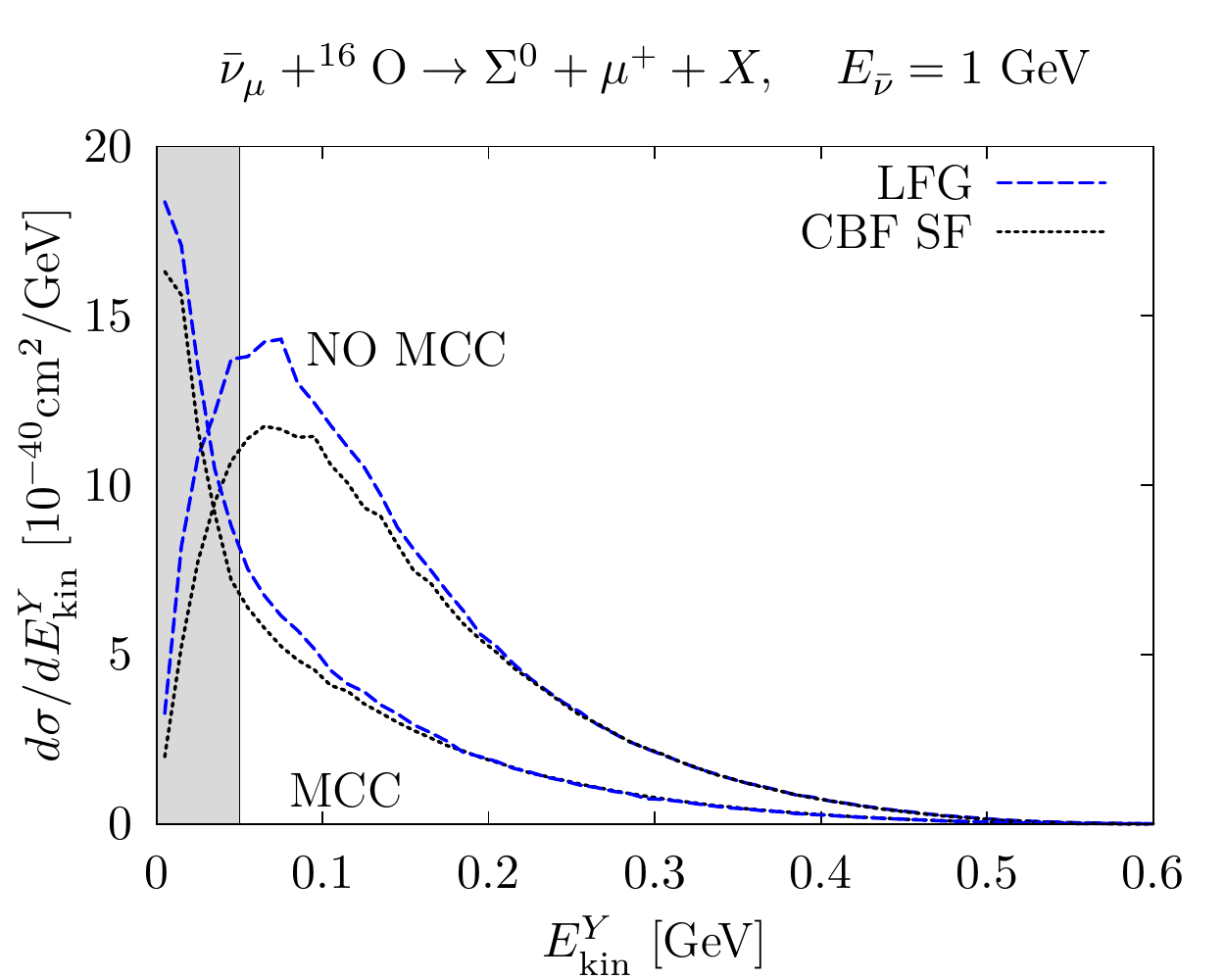}
\includegraphics[scale=0.7]{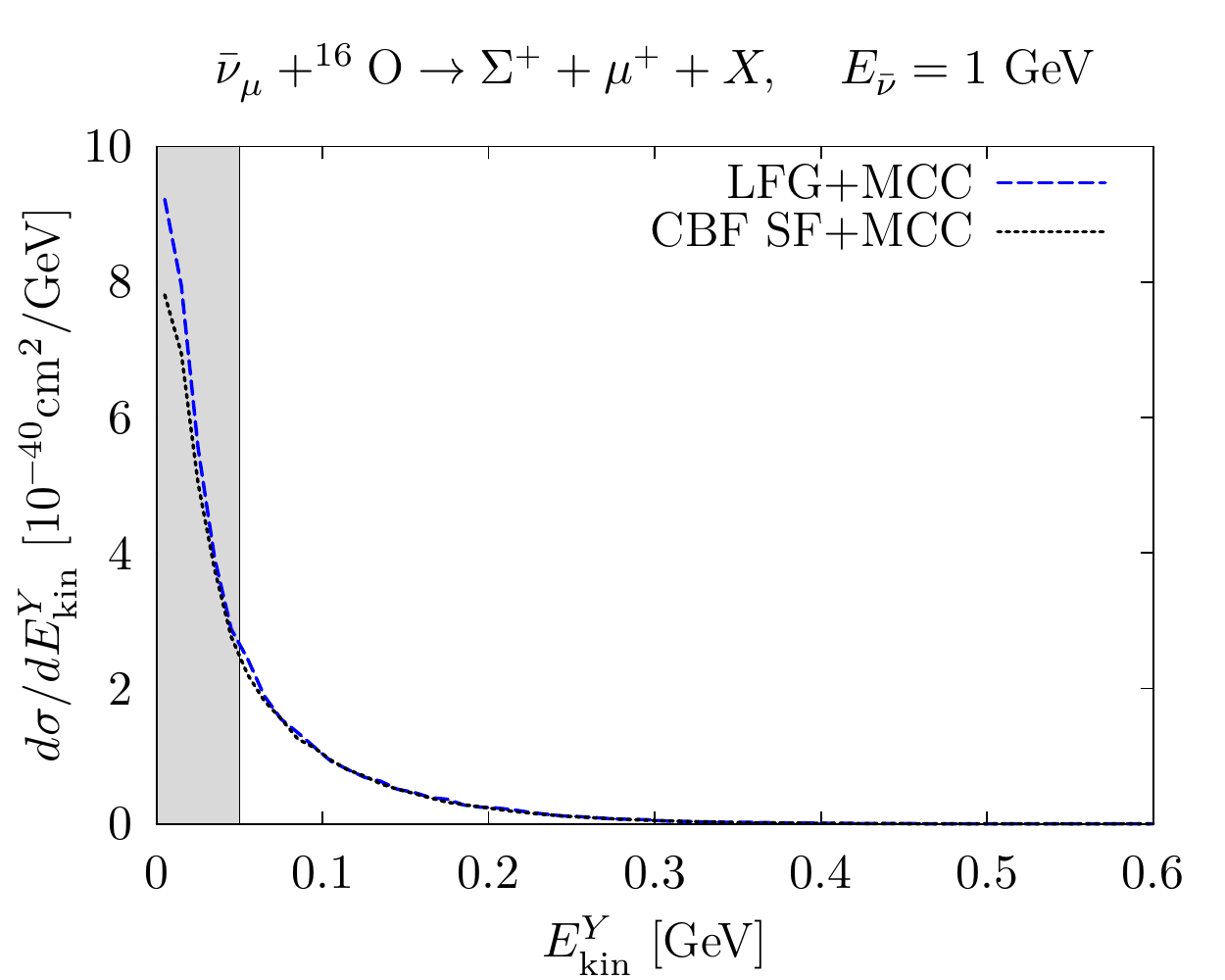}
\caption{Hyperon kinetic energy distributions, $d\sigma /dE^Y_{\rm kin}$,  for $\Lambda$, $\Sigma^-$, $\Sigma^0$, and $\Sigma^+$ production in oxygen and  $E_{\bar{\nu}}=1$ GeV. The different curves correspond to the LFG and CBF-SF predictions, with and without the inclusion of MCC effects. The mean-field potential corrections to the energy spectrum of the hyperon have not been included. The shaded areas correspond to $E^Y_{\rm kin} \leq 50$ MeV.  }
 \label{fig:dsigmadey}
\end{figure*}
%
\begin{figure*}[tbh]
\includegraphics[scale=0.7]{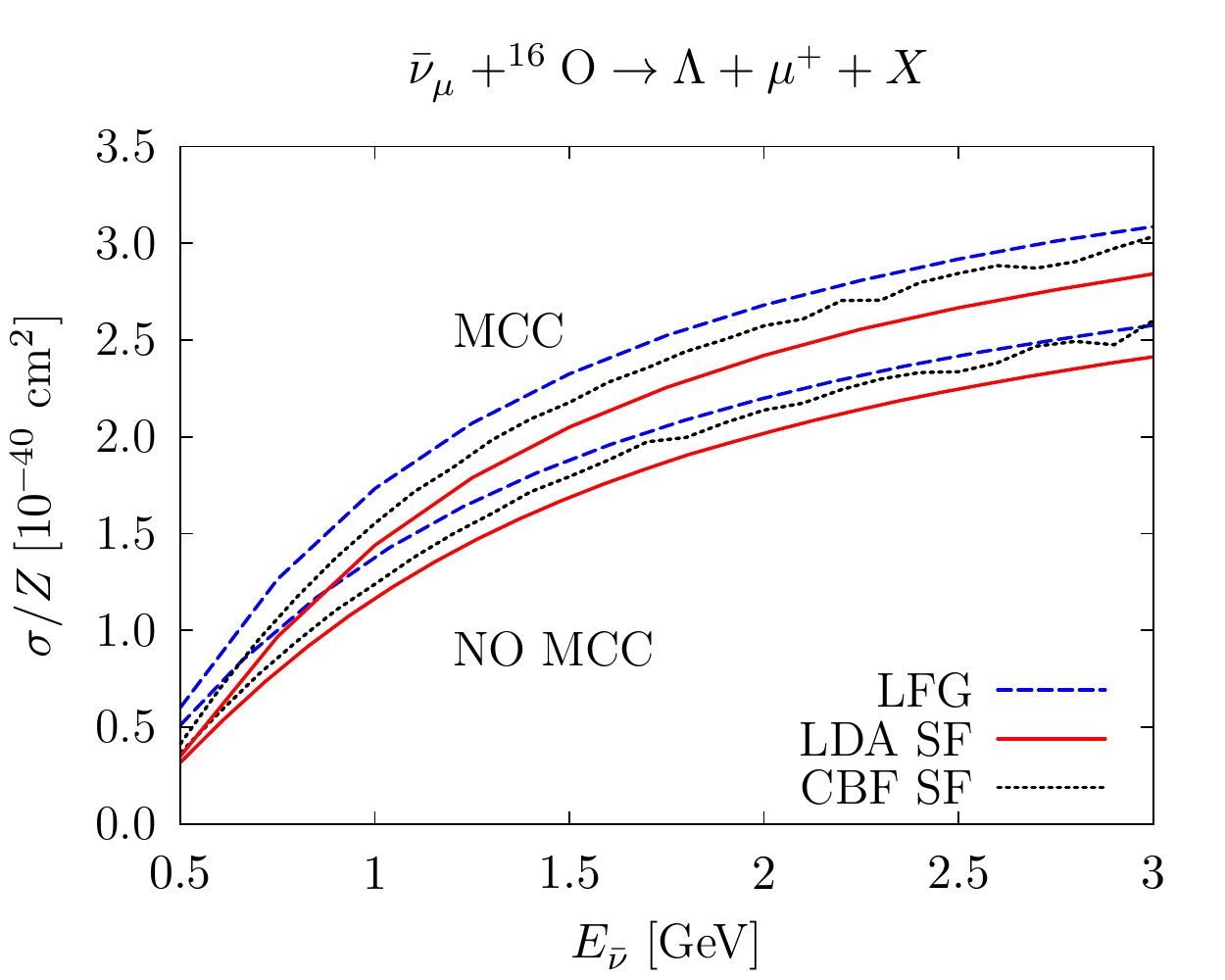}
\includegraphics[scale=0.7]{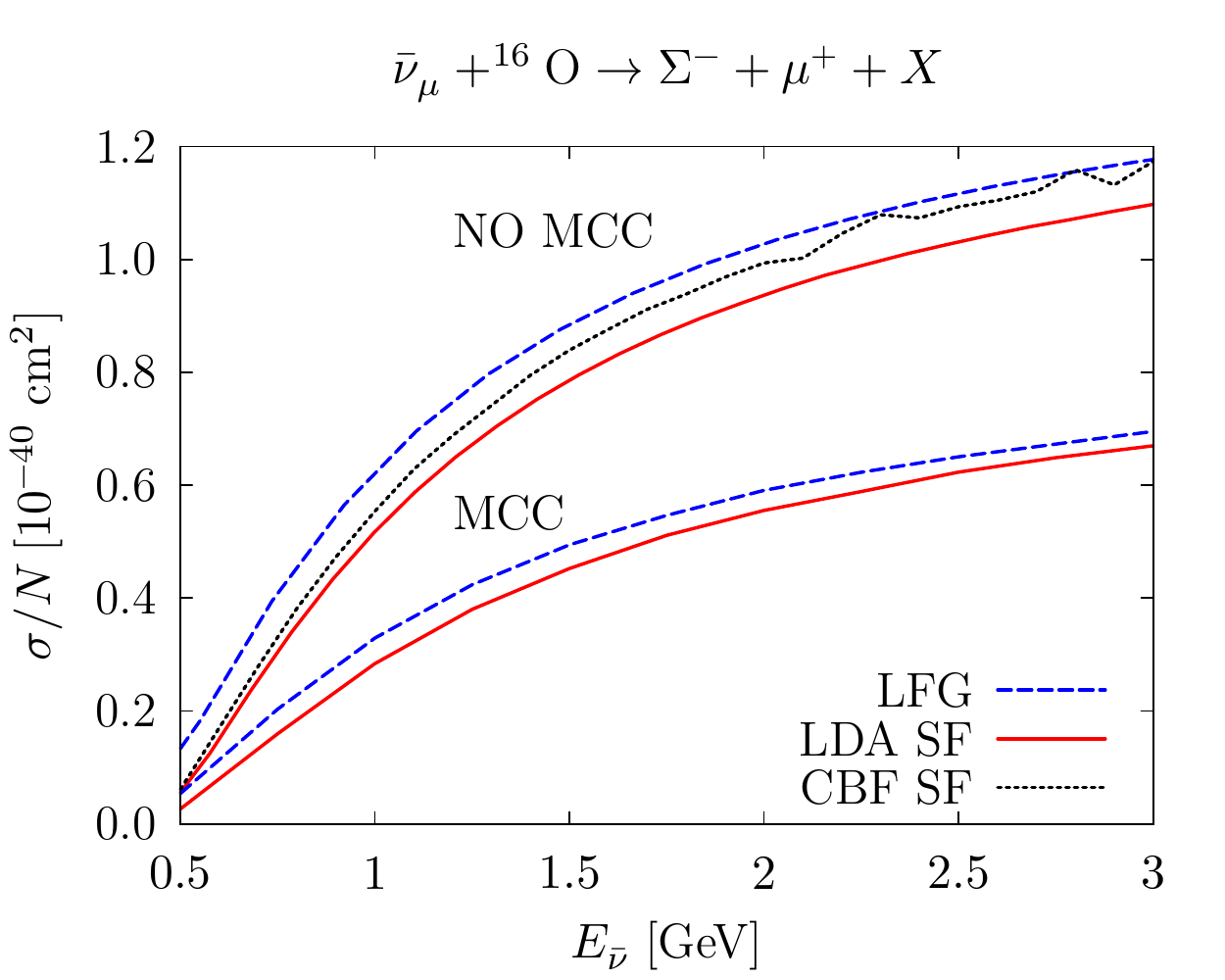}
\includegraphics[scale=0.7]{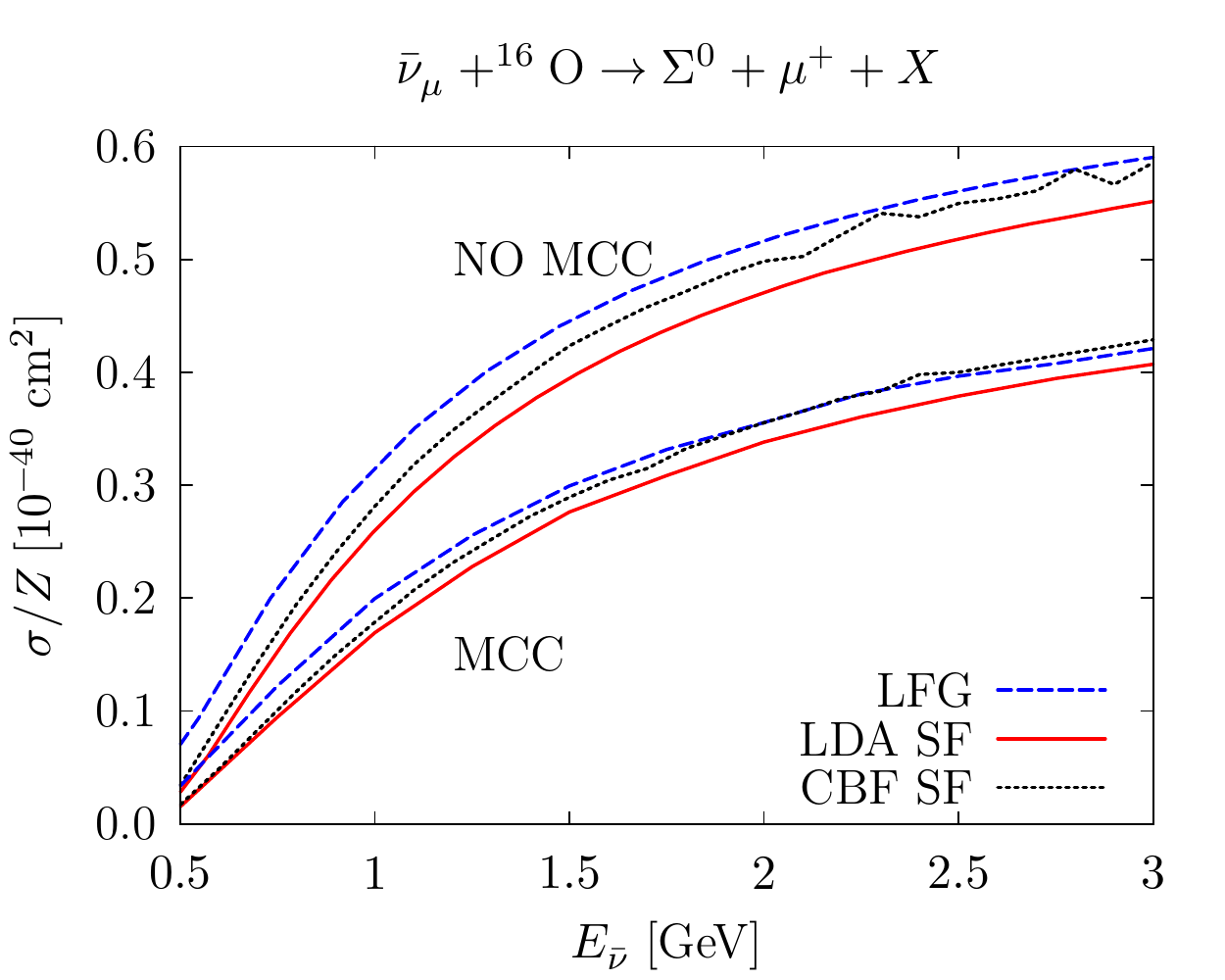}
\includegraphics[scale=0.7]{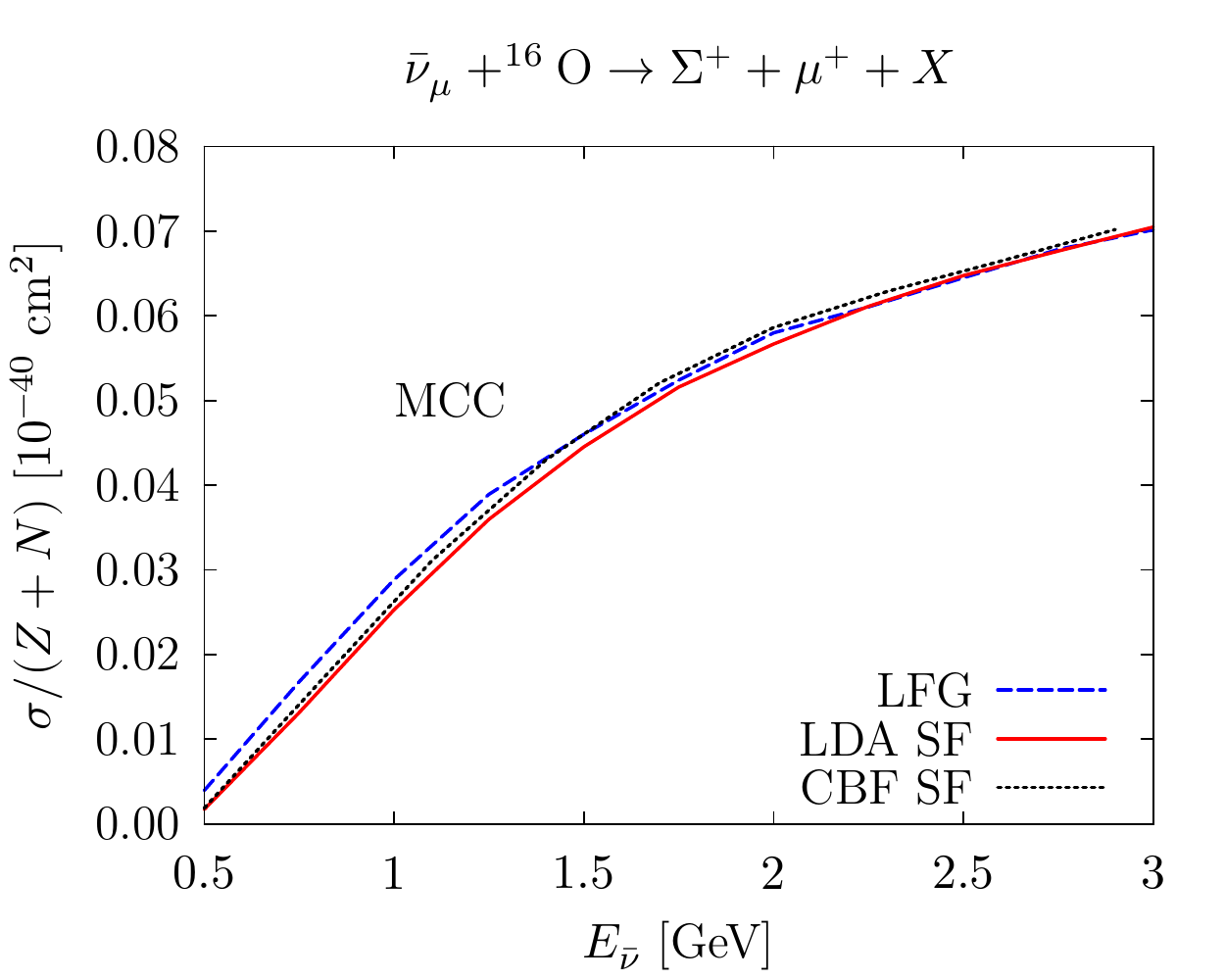}
\caption{Total cross sections for $\Lambda$, $\Sigma^-$, $\Sigma^0$ and $\Sigma^+$ production on $^{16}$O per active nucleon, as a function of the energy of the incoming $\bar{\nu}_\mu$. As for the $\Lambda$, $\Sigma^-$, and $\Sigma^0$ hyperons, we show results obtained with and without the MCC, while the $\Sigma^+$ hyperon can only be produced through secondary collisions.}
 \label{fig:sigma_tot}
\end{figure*}

In Fig.~\ref{fig:dsigmadey}, we display the $\Lambda$ and $\Sigma$ hyperon kinetic-energy distributions obtained using the LFG and the CBF-SF models. Similar results can be obtained using the LDA-SF. For a more direct comparison with Ref.~\cite{Singh:2006xp}, we have not included the hyperon MF potential. Although its effects are not negligible in the double-differential distributions displayed in Fig.~\ref{fig:dsigmadEdcosT}, they become very small in the single-differential and totally-integrated cross sections of Figs.~\ref{fig:dsigmadq2}, \ref{fig:dsigmadey} and \ref{fig:sigma_tot}. In the MCC, we used a threshold energy cut of 30 MeV for QE collisions. The shaded areas in Fig.~\ref{fig:dsigmadey} correspond to $E^Y_{\rm kin} \leq 50$ MeV; for such low values of the hyperon kinetic energy the details of the energy spectrum are not meaningful, although the integrals underneath the curves provide estimates of the total number of low-energy hyperons that are produced.
In analogy with Fig.~\ref{fig:dsigmadq2}, the inclusion of the MCC leads to sizable modifications of the initial differential cross sections. The $\Lambda$ channel is sizably enhanced at low $E^Y_{\rm kin}$ and depleted above $100$ MeV. The $\Sigma^-$ and $\Sigma^0$ cross sections are strongly quenched, except for very low $E^Y_{\rm kin}$ in the $\Sigma^0$ channel. Note, however, that our results are not much reliable in this region. Lastly, the $\Sigma^+$ production becomes non vanishing because of the secondary collisions accounted for the MCC. The LFG distributions displayed in Fig.~\ref{fig:dsigmadey} should be directly comparable to those shown in the left panels of Fig. 3 of Ref.~\cite{Singh:2006xp}. While the NO MCC  results nicely agree, we see that the effects of the MCC computed in this latter reference turn out to be much smaller and not as visible as those found in the present work. We have verified that this discrepancy has to be ascribed to a wrong implementation  in Ref.~\cite{Singh:2006xp} of the Pauli blocking of the outgoing nucleons produced in secondary collisions, leading to an important reduction of the number of interactions experienced by the knocked-out hyperons.

In Fig. \ref{fig:sigma_tot} we show the total cross sections for $\Lambda$, $\Sigma^-$, $\Sigma^0$ and $\Sigma^+$ production on oxygen. Consistently with the results of Fig.~\ref{fig:dsigmadq2}, there are small deviations among the curves referring to LDA-SF and CBF-SF results, which in general agree reasonably well. On the other hand, the application of the MCC noticeably modifies the theoretical predictions for all the hyperon channels; these effects being larger than those associated to the use of realistic hole SFs in place of the LFG model for the initial nuclear state. In absence of the MCC, the production rates of $\Sigma^-$ and $\Sigma^0$ are related by a SU(3) rotation. The associated Clebsh-Gordan coefficient, $\sqrt{2}$, leads to twice as large cross section for $\Sigma^-$ as for $\Sigma^0$. However, the corrections induced by the MCC alter this relation, and the reduction of the $\Sigma^-$ is about $20\%$ stronger than for $\Sigma^0$. 

\begin{figure}[tbh]
\includegraphics[scale=0.7]{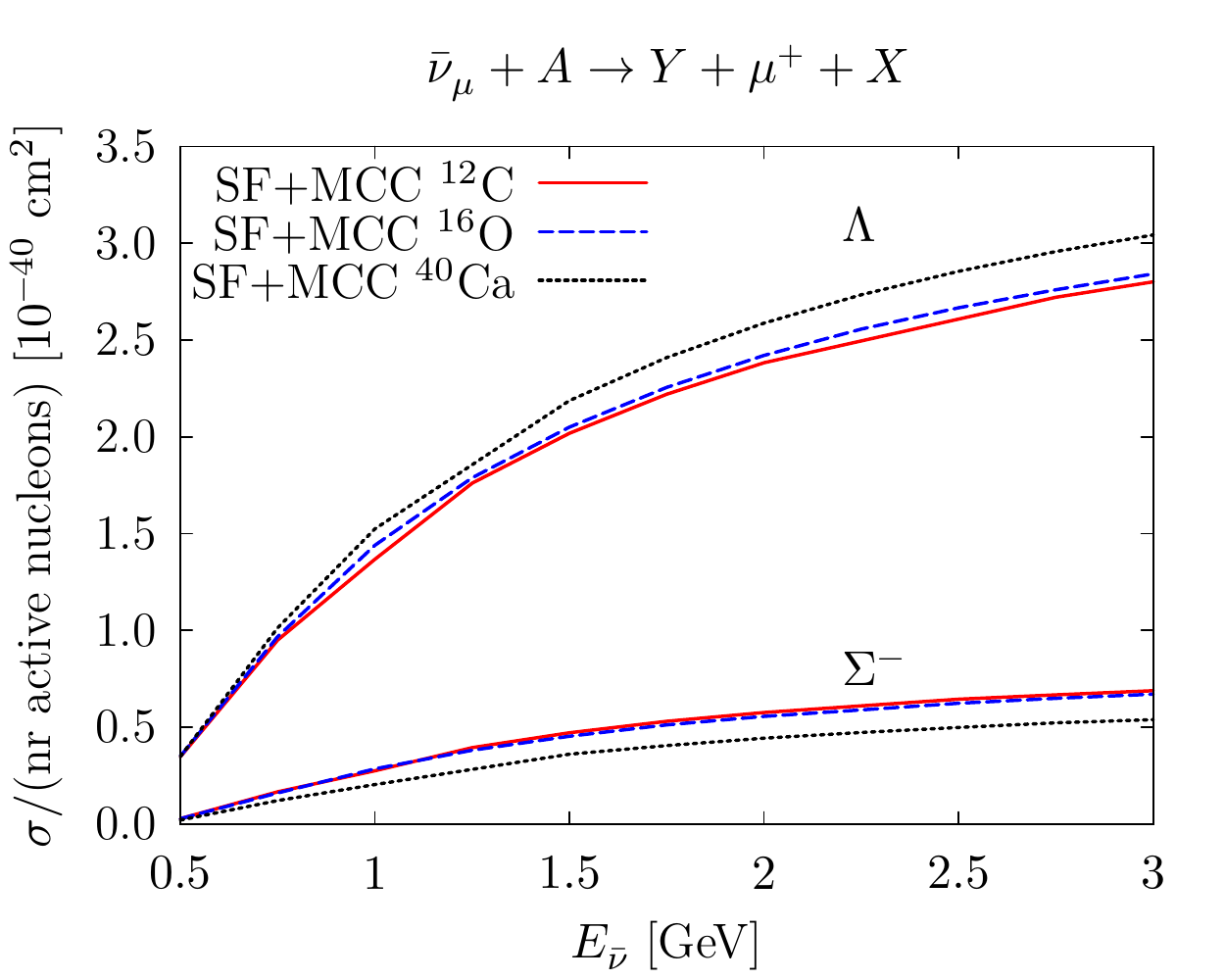}
\caption{Total $\Sigma^-$ and $\Lambda$ production  cross sections in $^{12}$C, $^{16}$O and $^{40}$Ca, as a function of the antineutrino energy. The results are obtained using the LDA-SF to model the initial nuclear state and the hyperon secondary collisions are accounted for through the MCC. The three lower (upper) curves correspond to the $\Sigma^-$ ($\Lambda$) production.}
 \label{fig:sigma_tot_nuclei}
\end{figure}

It is interesting to understand how the role played by the MCC depends upon the size of the nucleus. In this regard, in Fig. \ref{fig:sigma_tot_nuclei} we compare the total cross sections for $\Lambda$ and $\Sigma^-$ production on $^{12}$C, $^{16}$O and $^{40}$Ca, obtained using the LDA-SF. To make the comparison more transparent, for each nucleus we divided the total cross section by the corresponding number of active nucleons. While the $^{16}$O and $^{12}$C results are almost indistinguishable, there is around $10\%$ difference with those obtained for $^{40}$Ca. This is likely to be ascribed to the longer path that the hyperons have to travel before exiting the nucleus, implying a larger number of re-scattering processes.

\subsection{$\Lambda_c$ production}

In this subsection, we discuss  the results obtained for the $\nu_\mu+ ^{16}{\rm O}\rightarrow \mu^-+\Lambda_c+ X$ cross section using the different parameterizations of the $n\to \Lambda_c$ form factors discussed  in Subsec.~\ref{sec:lambdaC}. Our aim is to estimate how the large theoretical uncertainties of these form factors in the free space affect the predictions for the cross section of the $\Lambda_c$ production in nuclei. For simplicity, in this analysis, nuclear dynamical correlations in the initial state and hyperon FSI have been neglected altogether. Hence, we employ a simple LFG model to describe the initial nuclear target, we assume a free energy spectrum for the struck particle, and we do not employ the MCC. As for the latter, it has to be mentioned that there is no available experimental information on the $\Lambda_c$ mean free path in a nuclear environment, and the theoretical predictions for the interactions of the $\Lambda_c$ with nucleons and other charmed baryons suffer from severe uncertainties. 

\begin{figure}[tbh]
\includegraphics[scale=0.9]{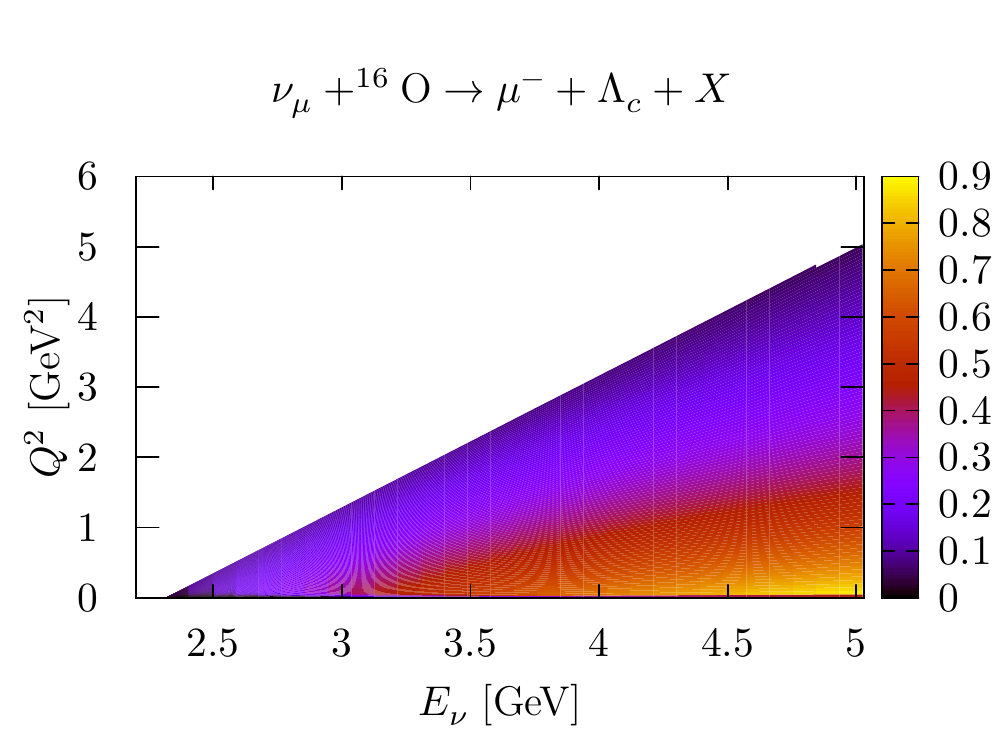}
\caption{Differential cross section $d\sigma(\nu_\mu+ ^{16}{\rm O}\rightarrow \mu^-+\Lambda_c^++ X)/dq^2$ [$10^{-40}\text{cm}^2\text{GeV}^{-2}$] per nucleon as a function of $E_\nu$.  The results are obtained within the LFG model for the initial nuclear state and assuming a free $\Lambda_c$ propagation in the final state. The weak-transition form factors are taken from the RCQM of Ref.~\cite{Gutsche:2014zna} and are extrapolated to the $Q^2>0$ region.}
 \label{fig:lambdaC_region}
\end{figure}

The energy distributions of the incoming neutrino fluxes of MINERvA \cite{Drakoulakos:2004gn} and DUNE \cite{Acciarri:2015uup} experiments peak at $E_\nu \simeq 3$ GeV and $E_\nu \simeq 5$ GeV, respectively. In Fig.~\ref{fig:lambdaC_region} we present the differential cross section $d\sigma/dq^2$ for the $\nu_\mu + ^{16}\text{O}\rightarrow \mu^-+\Lambda_c+X$ reaction, computed employing the RCQM form factors of Ref.~\cite{Gutsche:2014zna}, for incoming neutrino energies of up to 5 GeV. The maximum values of $Q^2$ reached in the production mechanism are $1.5$ GeV$^2$ and $5$ GeV$^2$, for $E_\nu=3$ GeV and $E_\nu=5$ GeV, respectively. However, the bulk of to the total cross section stems for $Q^2$ below $0.5$ GeV$^2$ and $2$ GeV$^2$, for neutrino energies corresponding to the peaks of the MINERvA and DUNE fluxes, respectively. These relatively low values of $Q^2$ justify the use of form factors fitted to the $\Lambda_c\to \Lambda$ semileptonic decay, corresponding to $Q^2\in [-1.36,0]$ GeV$^2$. 
%
However, extrapolating the form factors to moderately large positive $Q^2$ augment the theoretical uncertainty in our cross-section predictions. To estimate them, we have considered the five sets of form factors reviewed in Sec.~\ref{sec:lambdaC}, characterized by the different $q^2$ dependencies displayed in Fig.~\ref{fig:ff_charmNuc}.

\begin{figure*}[!htb]
\includegraphics[scale=0.7]{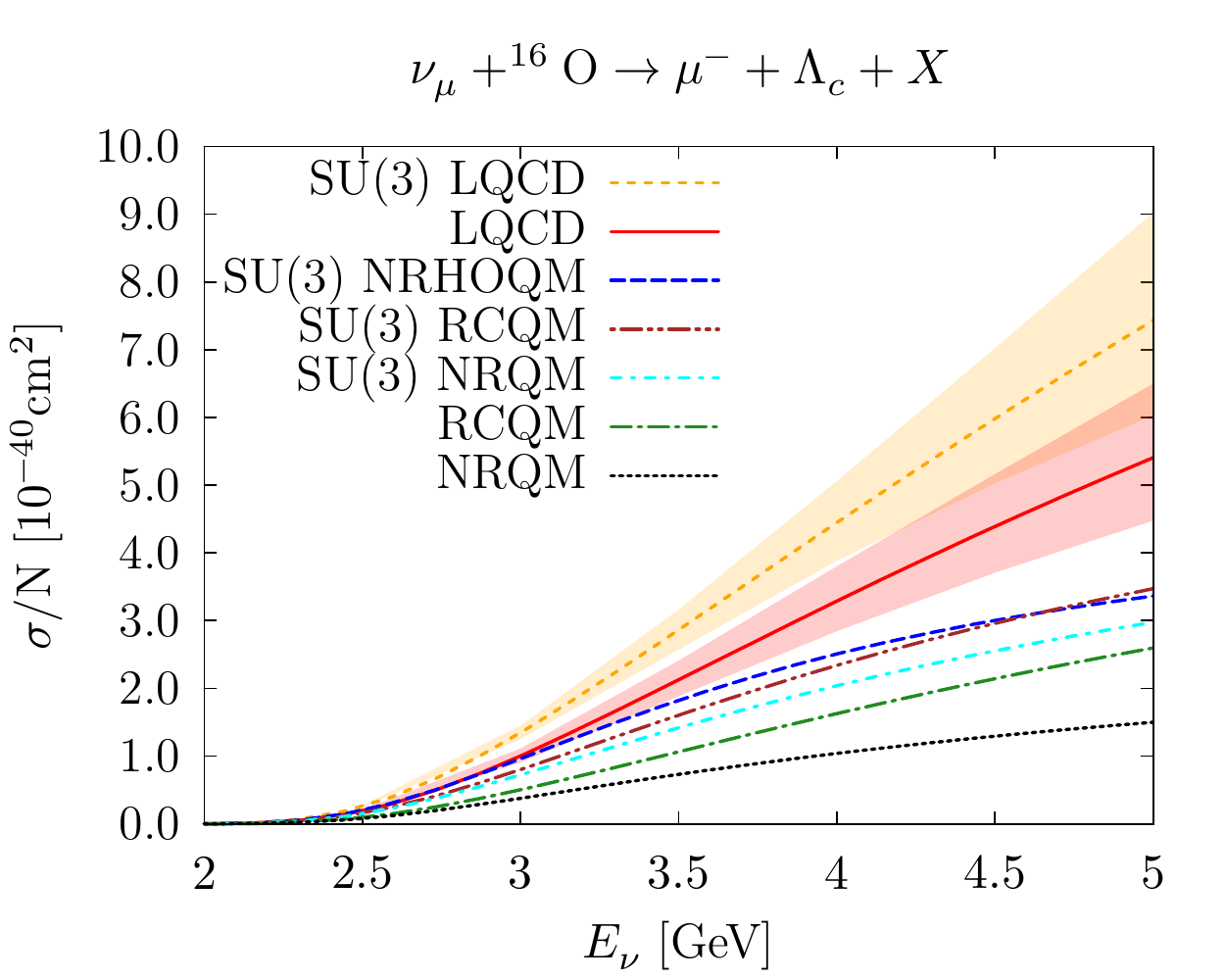}
\includegraphics[scale=0.7]{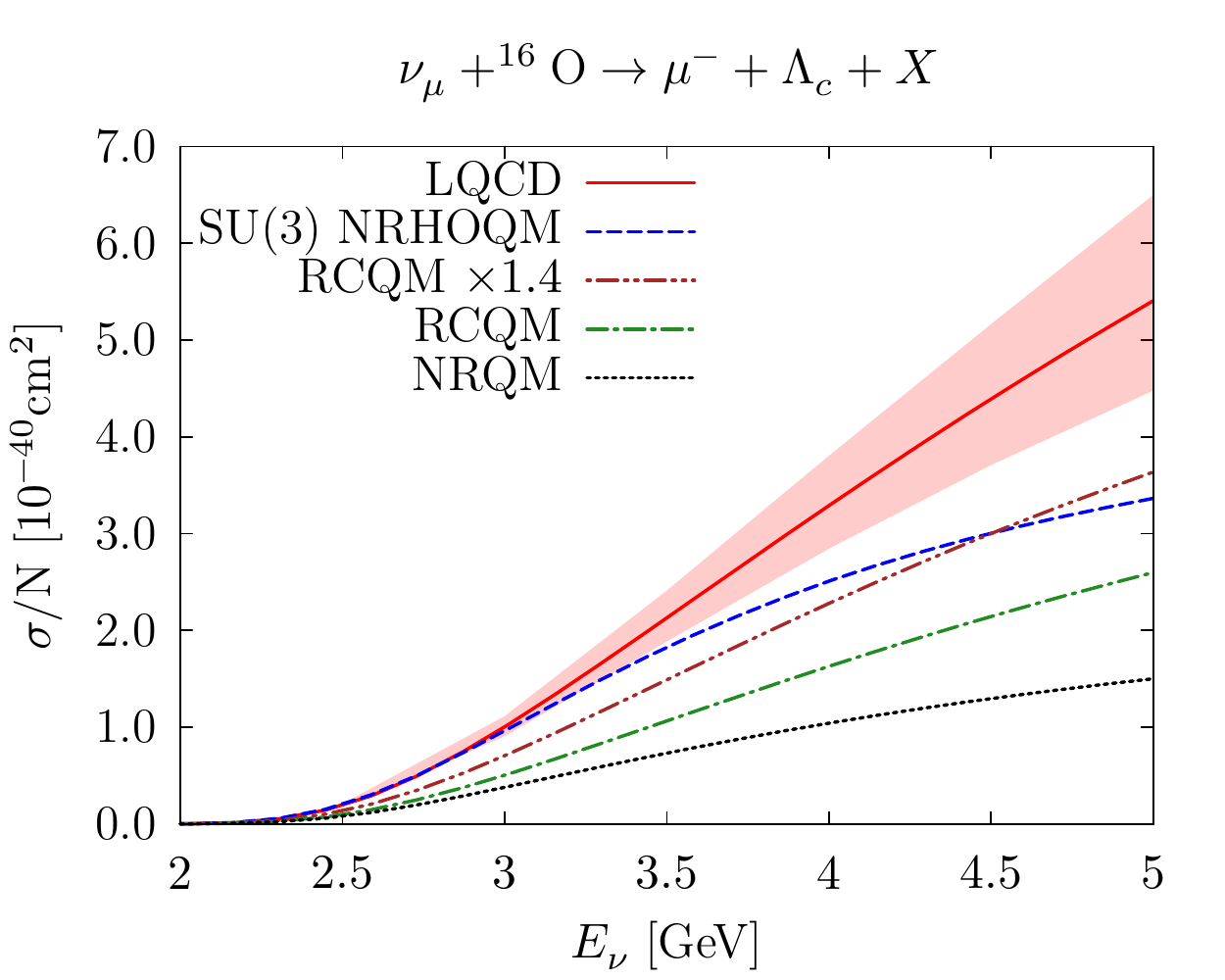}
\caption{Total cross section, per number of neutrons, for the $\nu_\mu + ^{16}\text{O}\rightarrow \mu^-+\Lambda_c^++X$ reaction, as a function of the incoming neutrino energy. The LFG model for the initial nuclear state is adopted and no FSI for the $\Lambda_c$ are accounted for. In the left panel the curves represent the set of weak transition form factors analyzed in the text: LQCD~\cite{Meinel:2017ggx,Meinel:2016dqj}, NRHOQM~\cite{Hussain:2017lir}, NRQM~\cite{PerezMarcial:1989yh}, and RCQM~\cite{Gutsche:2014zna,Gutsche:2015rrt}. The curves labeled as SU(3) represent the results attained multiplying the $\Lambda_c\rightarrow \Lambda$ form factors by the appropriate Clebsh-Gordan coefficient. The 68\% CL bands for the LQCD predictions are obtained from the Gaussian covariance matrices of Refs.~\cite{Meinel:2016dqj,Meinel:2017ggx}. In the right panel, we display results corresponding to the direct calculations of the $\Lambda_c\rightarrow N$ form factors, the only exception being the SU(3) NRHOQM predictions for which they are not available. The ``RCQM $\times 1.4$'' curve represents the results of Ref.~\cite{Gutsche:2014zna} rescaled by a factor 1.4, as inferred from the discussion on the $\Lambda_c\rightarrow \Lambda e^+\nu_e$ reported in the Appendix. }
 \label{fig:lambdaC}
\end{figure*}

As for the LQCD~\cite{Meinel:2017ggx}, the RCQM~\cite{Gutsche:2014zna}, and NRQM~\cite{PerezMarcial:1989yh} approaches\footnote{For the sake of clarity, we will not report results for the MBM model, as it does not add relevant information to our analysis. However, we include this model in the discussion of the $\Lambda_c\to \Lambda$ form factors carried out in the Appendix.}, in addition to the $\Lambda_c\rightarrow N$ form factors, we have also used those associated to the $\Lambda_c\rightarrow \Lambda$ transition, assuming unbroken SU(3) symmetry, and multiplying the matrix element by the appropriate Clebsh-Gordan coefficient ($\sqrt{3/2}$). When employing the latter form factors, the predicted  $\nu_\mu+ ^{16}{\rm O}\rightarrow \mu^-+\Lambda_c+ X$ cross sections turn out to be around  $10-30\%$ higher than those obtained from the direct weak $N \to \Lambda_c$ matrix element -- see the left panel of Fig.~\ref{fig:lambdaC} -- the smallest differences corresponding to the LQCD form factors. As for the NRHOQM of Ref.~\cite{Hussain:2017lir}, the results of Fig.~\ref{fig:lambdaC} have been obtained from the $\Lambda_c\rightarrow \Lambda$ form factors assuming SU(3) symmetry, although it might well be that this procedure overestimates the cross section. The 68\% CL uncertainty bands pertaining to the LQCD predictions are propagated from the errors and the covariance matrices of the fit parameters listed in Refs.~\cite{Meinel:2016dqj,Meinel:2017ggx}. These theoretical uncertainties grow with the neutrino energy, and become as large as $\sim 20\%$ for $E_\nu=5$ GeV. 

There are sizable discrepancies in the $\nu_\mu+ ^{16}{\rm O}\rightarrow \mu^-+\Lambda_c+ X$ cross sections corresponding to the various form factors, despite that all of them are constrained by the experimental $\Lambda_c\rightarrow \Lambda e^+\nu_e$ decay width. The reason for this behavior is twofold. On the one side, the form factors exhibit major differences in their $Q^2$ dependence. For instance, those from the NRHOQM ~\cite{Hussain:2017lir} are suppressed with increasing $Q^2$, leading to reduced cross sections, when compared to other approaches -- see the detailed discussion of Figs.~\ref{fig:lambdaC_decay} and \ref{fig:ff_charmLambda} in the Appendix. On the other side, the relative importance of the vector and axial terms of the current  to the decay width turns out to be strongly model dependent. The vector contribution amounts to only $\sim10\%$ of the total $\Lambda_c \to \Lambda $ semileptonic width for the NRQM, it becomes around $30\%$ for the NRHOQM and RCQM, and it is about $\sim 34\%$ in the case of the LQCD approach. The vector form factors, $f_1$ in particular, play a more important role in the determination of the $\Lambda_c$ neutrino-production cross section. Hence, it is not surprising that NRQM model, predicting a relatively small vector form factors, leads to the lowest cross-section estimates. 

In the right panel of Fig.~\ref{fig:lambdaC}, we display the cross sections corresponding to the direct calculations of the $\Lambda_c\rightarrow N$ form factors within the different approaches. The NRHOQM predictions, for which this option is not available, are rescaled assuming SU(3) symmetry. We also, show the results of RCQM of Ref.~\cite{Gutsche:2014zna}, rescaled by a factor 1.4, as inferred from the discussion on the $\Lambda_c\rightarrow \Lambda e^+\nu_e$ total and differential widths carried out and represented in Fig.~\ref{fig:lambdaC_decay} of the Appendix. This factor amounts for the difference in the total semileptonic decay width between the RCQM and the LQCD predictions reported in Refs.~\cite{Gutsche:2015rrt} and \cite{Meinel:2016dqj}, respectively. Note that both approaches provide similar $Q^2$ dependencies of the leading form factors, $f_1$ and $g_1$, not only in the semileptonic decay phase-space, but also for the positive $Q^2-$values, relevant for the neutrino-production reaction -- see the discussions of Figs.~\ref{fig:ff_charmNuc} of Subsec.~\ref{sec:lambdaC} and \ref{fig:ff_charmLambda} of the Appendix. 

The total $\nu_\mu+^{16}\text{O} \rightarrow \mu^-+ \Lambda_c +X$ cross section can be reasonably estimated to lie between the LQCD and the rescaled RCQM predictions. This range also accommodates the SU(3) NRHOQM results, and it leads to theoretical uncertainties below 30\% up to $E_\nu \sim 3.5$ GeV. For neutrino energies corresponding to the peaks of the MINERvA and DUNE fluxes, we thus estimate the total cross sections to be $\sigma(E_\nu=3\,\text{GeV})/N=(0.9^{+0.2}_{-0.1})\times 10^{-40}\text{cm}^2$ and $\sigma(E_\nu=5\,\text{GeV})/N=(4.5^{+2.0}_{-0.9})\times 10^{-40}\text{cm}^2$, respectively. The central value is the average between the LQCD and the rescaled RCQM results, while the errors account for the difference of these two sets of results, taking into account the upper limit of the LQCD uncertainty band. Taking a somewhat less conservative perspective, if the LQCD predictions are assumed to be sufficiently reliable, the corresponding estimates for the cross sections can be readily inferred the plots. Nevertheless, one should also bear in mind that $\Lambda_c$ FSI effects have not been considered in this preliminary analysis. Although their role is expected to be less relevant than for strange hyperons, they will have some impact in the production rate of $\Lambda_c$ baryons.

\section{Conclusions}\label{sec:concl}

We have studied the weak production of $\Lambda$ and $\Sigma$ hyperons induced by $\bar{\nu}_\mu$ scattering off nuclei, carefully considering the effects of nuclear dynamics. To describe correlations in the initial nuclear target, we employed hole SFs obtained within two many-body methods, both successfully applied to study electroweak reactions in nuclei. The propagation of the hyperons in the nuclear medium is tackled by a MCC algorithm, which treats the rescattering processes in a classical fashion. For this reason, the MCC does not modify the inclusive ($\bar\nu_l,l^+ Y)$ cross sections, if the sum for $Y=\Lambda, \Sigma^0$ and $\Sigma^\pm$ is considered. On the other hand, when more exclusive processes are analyzed, such as the production rates, the energies, and the angular distributions of the specific hyperons' species, the MCC plays a major role. For instance, although the $\Sigma^+$ hyperon is not produced in the interaction vertex, because of secondary collisions its production rate  in the scattering process does not vanish. Of course, when all the possible channels are summed, the inclusive cross section is recovered. 

We find that our calculations carried out employing semi-phenomenological LDA-SF~\cite{FernandezdeCordoba:1991wf} and the CBF-SF~\cite{Benhar:1989aw,Benhar:1994hw} nicely agree, providing robust estimates of the importance of dynamical correlations in the initial nuclear state, neglected in the pioneering work of Ref.~\cite{Singh:2006xp}, based on the LFG model. We show how the inclusive double-differential $d^2\sigma/(d\cos\theta' dq^0)$ distributions are significantly affected, as the correlations encoded in the hole-SFs bring about a reduction of the height of the QE peak and a redistribution of the strength to higher energy-transfers regions. In the case of the differential cross section $d\sigma/dq^2$ and the total one, the effects of considering realistic hole-SFs are not as important as in the $d^2\sigma/(d\cos\theta' dq^0)$ case and are less relevant than those produced by the FSI of the hyperon. The MCC strongly modifies the impulse-approximation results for the exclusive processes, leading to a non-zero $\Sigma^+$ cross section, to a sizable enhancement of the $\Lambda$ production and to a drastic reduction of the $\Sigma^0$ and $\Sigma^-$ distributions.  

It has to be mentioned that the MCC effects found in this work are much larger and more visible than those reported in Ref.~\cite{Singh:2006xp}. This is due to an incorrect implementation of the Pauli blocking of the outgoing nucleons produced in secondary collisions in the latter work, that led to an important reduction of the number of secondary collisions experienced by the hyperons during their path through the nucleus.

Motivated by the recent BESIII measurements of the branching ratios of $\Lambda_c\rightarrow \Lambda l^+\nu_l$ ($l=e,\mu$) decays and by the CHORUS results for the ratio of the cross section for $\Lambda_c$ production in neutrino-nucleon (CC) scattering, we computed the QE weak $\Lambda_c$ production cross section on nuclei. We have paid special attention to estimate the impact of the  $n\to \Lambda_c$ matrix-element theoretical uncertainties. To this aim, we employed form factors computed within different approaches ranging from the LQCD calculations of Ref.~\cite{Meinel:2017ggx} to state-of-the-art nonrelativistic and relativistic quark models of Refs.~\cite{PerezMarcial:1989yh,Gutsche:2014zna,Gutsche:2015rrt,Hussain:2017lir}. We find significant variations in the predictions from the different schemes adopted to compute the relevant form factors, despite the fact that all of them are constrained by the experimental $\Lambda_c\rightarrow \Lambda e^+\nu_e$  decay width. This is a direct consequence of the unavoidable ambiguities induced by extrapolating the form factors from the $q^2$ region relevant for the $\Lambda_c$ decay to the kinematics relevant to the $\Lambda_c$ production. The theoretical uncertainties are estimated to be below $30\%$ for $E_\nu \lesssim 3.5$ GeV. For the neutrino energies corresponding to the peaks of the MINERvA and DUNE fluxes, we predict the cross sections -- normalized to the number of neutrons -- to be $\sigma(E_\nu=3\,\text{GeV})/N=(0.9^{+0.2}_{-0.1})\times 10^{-40}\text{cm}^2$ and  $\sigma(E_\nu=5\,\text{GeV})/N=(4.5^{+2.0}_{-0.9})\times 10^{-40}\text{cm}^2$, respectively. For simplicity, in this preliminary analysis, dynamical correlations in the initial state and FSI effects have been neglected. 
 
\begin{acknowledgments}
 We warmly thank M. Vicente-Vacas, L. \'Alvarez-Ruso, J.T. Sobczyk, and A. Ankowski for useful discussions. This research  has been supported by the Spanish Ministerio de Ciencia, Innovaci\'on  y Universidades and European FEDER funds under  Contracts FIS2017-84038-C2-1-P and SEV-2014-0398. This work is supported by the U.S. DOE, Office of Science, Office of Nuclear Physics, under contract DE-AC02-06CH11357 (A. L. and N. R.), by Fermi Research Alliance, LLC, under Contract No. DE-AC02-07CH11359 with the U.S. Department of Energy, Office of Science, Office of High Energy Physics (N. R.). Numerical calculations have been
made possible through a CINECA-INFN agreement, providing access to resources on MARCONI at CINECA. 
 \end{acknowledgments}

\appendix
\section{Further details on the $\Lambda_c\to \Lambda, N$ form factors}\label{sec:appendix}
%

\begin{figure*}[htb]
\includegraphics[scale=0.9]{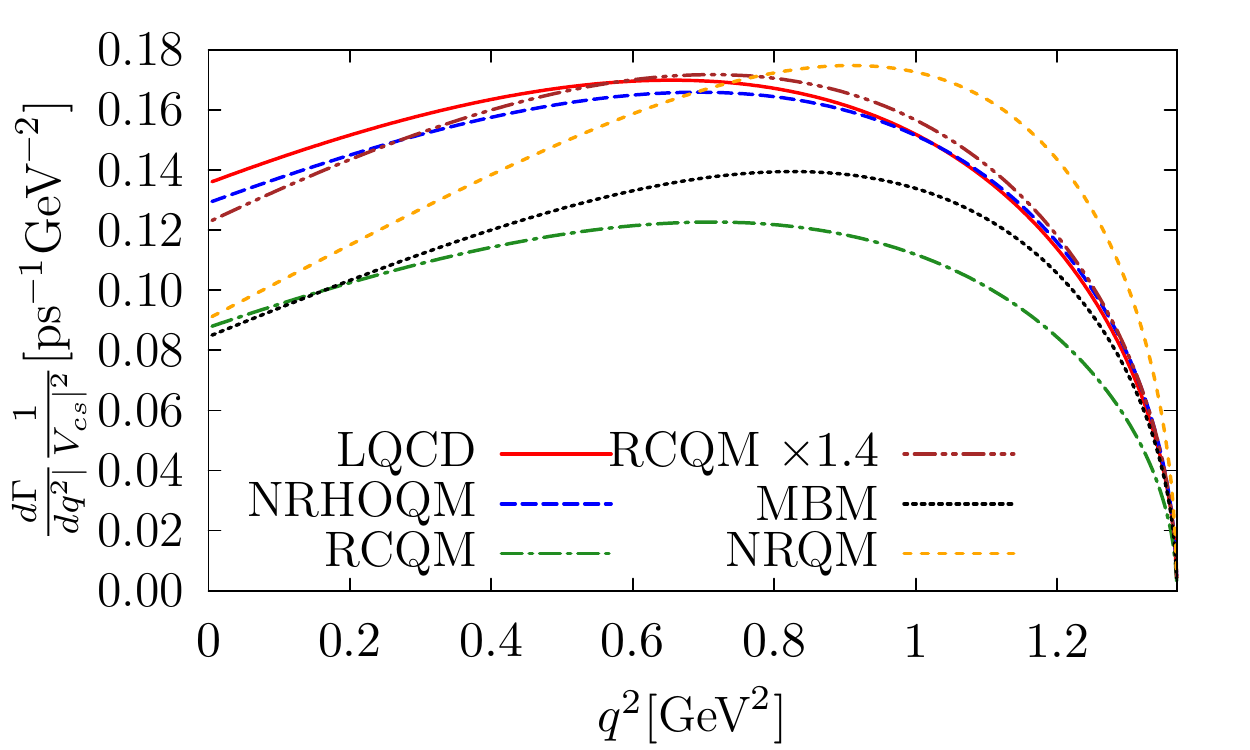}
\caption{ $\Lambda_c\rightarrow \Lambda e^+\nu_e$  differential decay width from NRHOQM~\cite{Hussain:2017lir}, RCQM~\cite{Gutsche:2015rrt},  MBM and NRQM~\cite{PerezMarcial:1989yh}, CQM approaches, and the LQCD simulation of Ref.~\cite{Meinel:2016dqj}.}
 \label{fig:lambdaC_decay}
\end{figure*}

\begin{figure*}[hbt]
\includegraphics[scale=0.8]{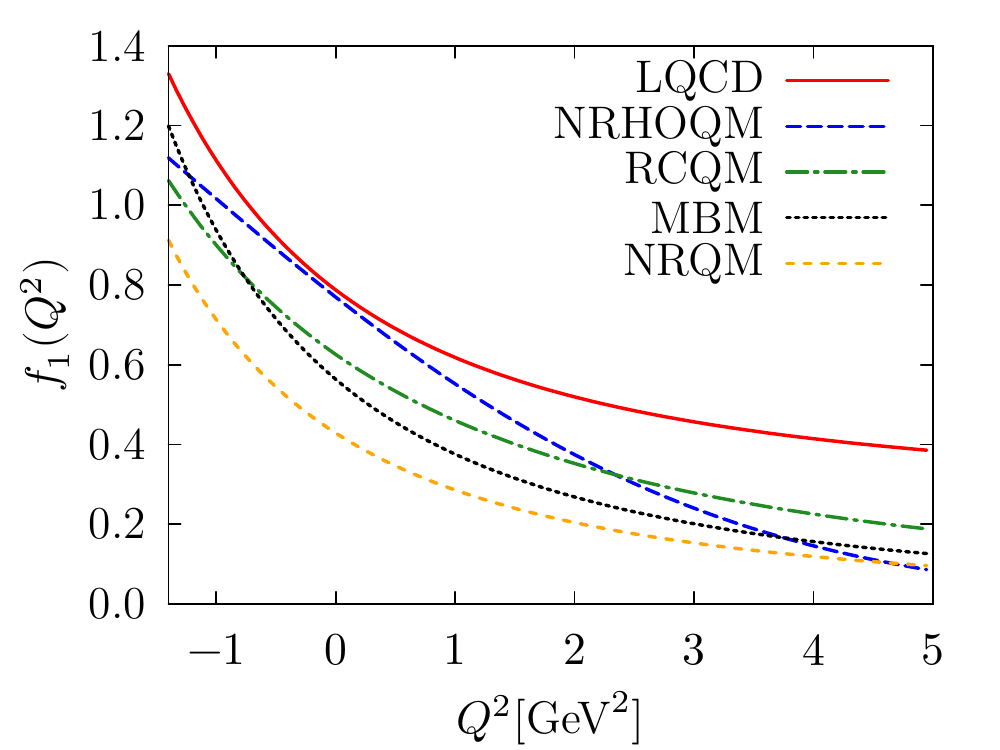}
\includegraphics[scale=0.8]{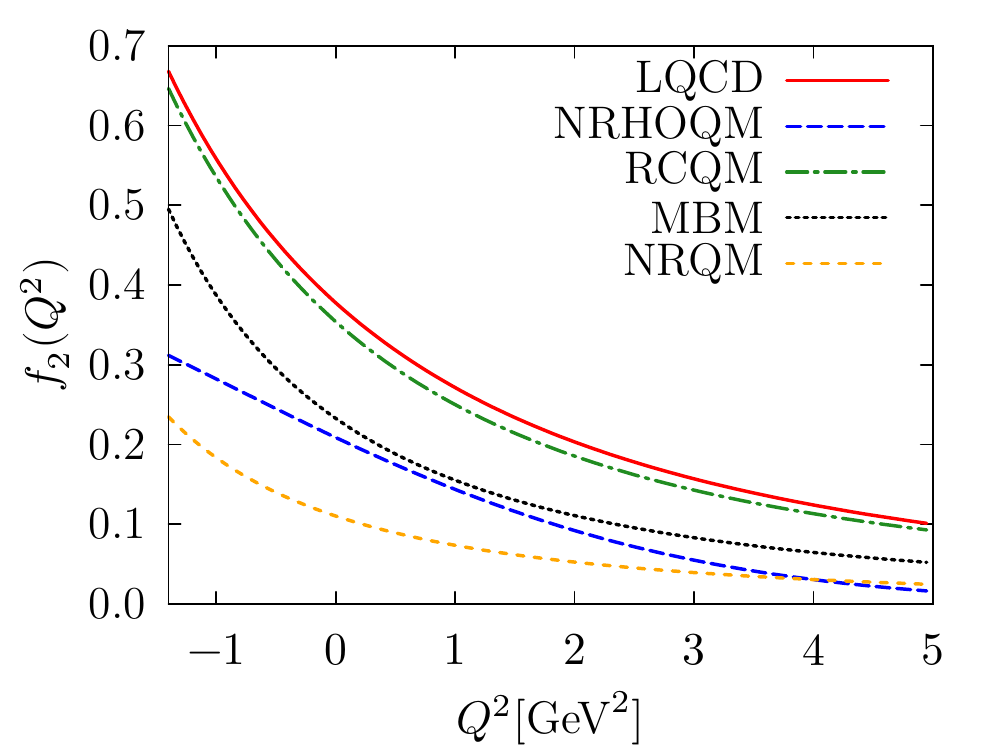}
\includegraphics[scale=0.8]{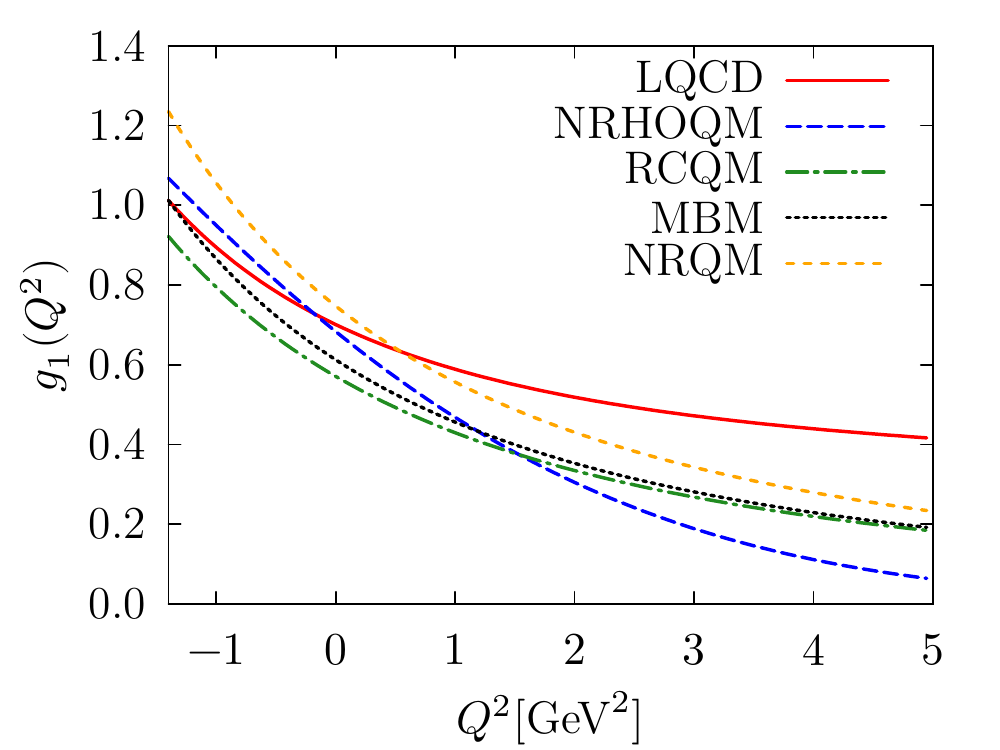}
\includegraphics[scale=0.8]{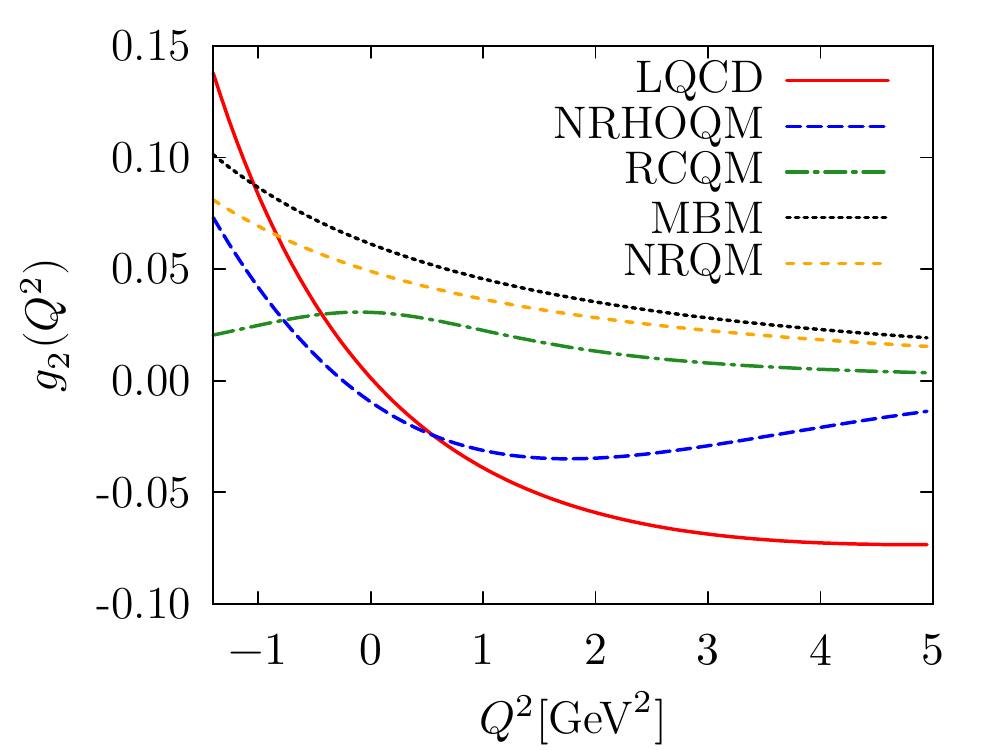}
\caption{Vector and axial form factors for the $\Lambda_c\rightarrow \Lambda$ transition calculated for the different models as detailed in Sec.~\ref{sec:lambdaC}. The form factors are multiplied by the Clebsh-Gordan coefficient $\sqrt{3/2}$ in order to estimate them for the $n\to \Lambda_c $ transition, assuming unbroken SU(3) flavor symmetry. Negative values of $Q^2$ correspond to the kinematics of the $\Lambda_c$ semileptonic decay, while $Q^2$ is positive for the $\Lambda_c$ neutrino production.}
 \label{fig:ff_charmLambda}
\end{figure*}

When computing the $\nu_\mu + n \rightarrow \mu^- + \Lambda_c^+$ cross section, we might assume SU(3) symmetry to relate the $\Lambda_c\rightarrow \Lambda$ form factors to the $\Lambda_c\rightarrow N$ ones. To estimate SU(3)-breaking effects, we have computed the $\mathcal{BR}(\Lambda_c\rightarrow \Lambda e^+\nu_e)$ comparing the results obtained using either the rescaled $\Lambda_c\rightarrow N$ or the $\Lambda_c\rightarrow \Lambda$ form factors. We performed this comparison for the LQCD, RCQM,  NRQM and MBM approaches of Refs.~\cite{Meinel:2016dqj,Gutsche:2015rrt, PerezMarcial:1989yh}. As for the LQCD form factors, the branching ratios attained assuming SU(3) symmetry are $\sim 20\%$ lower than the direct calculations with the actual $\Lambda\rightarrow \Lambda_c$ form factors. This reduction is even more evident, $\sim 40-50\%$, for the RCQM, NRQM, and MBM approaches. This source of uncertainties should be considered when analyzing the different predictions for $\sigma (\nu_\mu + ^{16}\text{O}\rightarrow \mu^-+\Lambda_c+X)/N$ displayed in Fig.~\ref{fig:lambdaC}.

The differential widths $d\Gamma(\Lambda_c\rightarrow \Lambda e^+\nu_e)/dq^2$ are shown in Fig.~\ref{fig:lambdaC_decay} for all the approaches used to compute the $\Lambda\rightarrow \Lambda_c$ form factors. LQCD, NRHOQM and MBM provide similar distributions, while the NRQM yields significantly different shape, the strength being shifted towards higher values of $q^2$. Note however, that no major differences in the total width are observed. As discussed in Subsec.~\ref{sec:lambdaC}, the RCQM predicts the lowest branching ratio, about $\sim 40\%$ smaller than the LQCD result reported in Ref.~\cite{Meinel:2016dqj}. However, once the RCQM differential distribution~\cite{Gutsche:2015rrt} is rescaled by a factor $1.4$, the shape of $d\Gamma/d q^2$ turns out to be very close to the LQCD one.

The main contribution to the total width ($\sim65\%$ in most of the models) comes from the axial form factor $g_1(Q^2)$, while the rest of the strength is driven by $f_1(Q^2)$ -- there is no axial-vector interference contributing to $d\Gamma/dq^2$. Because of the kinematical reasons, the ratio of the vector to axial parts is different for the $\nu_\mu + n \rightarrow \mu^- + \Lambda_c$ reaction, with $f_1(Q^2)$ playing a more important role. Because both $f_1$ and $g_1$ dominate the total cross section, it is interesting to pay some further attention to their behavior, when extrapolating them to the region of $Q^2$ relevant for neutrino scattering. The $Q^2$ dependence of the $\Lambda_c \to \Lambda$ form factors is displayed in Fig.~\ref{fig:ff_charmLambda}. They are multiplied by the factor $\sqrt{3/2}$, as dictated by the SU(3) symmetry to get the $c \to d$ matrix elements from the $c \to s$ ones. Besides the phase-space available in the semileptonic decay, corresponding to $Q^2 <0$, we also show the behavior of the form factors in the $Q^2 >0$ region accessible in $\Lambda_c$ neutrino-production reactions. It is interesting to compare the LQCD, RCQM, NRQM and MBM form factors of Fig.~\ref{fig:ff_charmLambda} with those shown in Fig.~\ref{fig:ff_charmNuc}, the latter being computed employing the appropriate $\Lambda_c\to N$ form factors. Such comparison helps estimating the size of SU(3)-breaking contributions, complementing the discussion on integrated $\mathcal{BR}(\Lambda_c\rightarrow \Lambda e^+\nu_e)$ that we alluded to earlier. Focusing on $f_1$ and $g_1$, these SU(3)-breaking effects are more apparent in the vicinity of $Q^2=0$, while they are less important as $Q^2$ increases, becoming moderately small for $Q^2=5$ GeV$^2$.

Fig.~\ref{fig:ff_charmLambda} shows that the LQCD and RCQM calculations for $f_1(Q^2)$, $f_2(Q^2)$ and $g_1(Q^2)$ exhibit a similar $Q^2$ dependence, while their predictions for $g_2(Q^2)$ are quite different. Note, however, that this is not much relevant for the neutrino-induced $\Lambda_c$ production, since the $g_2$ contributions for this process are negligible. As for the absolute size, both approaches predict almost the same $f_2(Q^2)$, but the RCQM results for $f_1(Q^2)$ and $g_1(Q^2)$ at $Q^2<0$ are $\sim 20\%$ smaller than those of LQCD. This discrepancy is more significant in the region of large $Q^2$, relevant for scattering processes. This latter behavior is less visible in the $\Lambda_c\rightarrow N$ transition form factors shown in Fig. \ref{fig:ff_charmNuc}, where the LQCD and RCQM shapes of $f_1$ and $g_1$ turn out to be remarkably similar for the entire range of $Q^2$, translating into the comparable $d\Gamma/d q^2$ shapes of Fig.~\ref{fig:lambdaC_decay}. 

We should also note that the $Q^2$ dependencies of the NRHOQM form factors are significantly different than those of the other models considered in this work. By only looking at the $d\Gamma/dq^2$ shape in Fig.~\ref{fig:lambdaC_decay}, one might expect that the NRHOQM and the LQCD predictions for the neutrino-induced $\Lambda_c$  production to be similar. However, this is not the case, as shown in Fig.~\ref{fig:lambdaC}, where the NRHOQM results lie well below those from LQCD. This is because $f_1$ and $g_1$ computed within the NRHOQM becomes smaller with increasing $Q^2$ much more rapidly than in any of the other models. Finally, both the NRQM and the MBM predict $g_1$ comparable with those obtained within other schemes. However, their results for $f_1$ at $Q^2>0$ lie below the results of LQCD, RCQM, and NRHOQM. Nevertheless, the NRQM and MBM $\Lambda_c \to \Lambda$ form factors are in an overall better agreement with those of the RCQM than in the case of the $\Lambda_c \to N$ transition shown in Fig.\ref{fig:ff_charmNuc}.

\bibliography{biblio}

\end{document}